\documentclass[twocolumn,aps,showpacs,prb,tightenlines,amsmath,amssymb]{revtex4}
\usepackage{graphicx}
\usepackage{amssymb}
\usepackage{dcolumn} 
\usepackage{amsmath}
\usepackage{bm}
\usepackage{colordvi}
\usepackage{mathrsfs}
\newcommand{\bgreek}[1]{\mbox{\boldmath$#1$\unboldmath}}
\makeatletter

\newcommand{\Rmnum}[1]{\expandafter\@slowromancap\romannumeral #1@}
\makeatother

\begin{document}

\title{Microscopic theory of ultrafast dynamics of carriers photoexcited by THz
and near-infrared linearly-polarized laser pulses in graphene} 
\author{B. Y. Sun} 
\author{M. W. Wu}
\thanks{Author to  whom correspondence should be addressed}
\email{mwwu@ustc.edu.cn.}
\affiliation{Hefei National Laboratory for Physical Sciences at
  Microscale and Department of Physics,
University of Science and Technology of China, Hefei,
  Anhui, 230026, China}
\date{\today}

\begin{abstract}
We investigate the dynamics of photoexcited carriers and nonequilibrium phonons
in graphene by solving the microscopic kinetic Bloch equations. The pump and drift effects
from the laser field as well as the relevant
scatterings (including the Coulomb scattering with dynamic screening) are
explicitly included. When the pump-photon energy is 
high enough, the influence of the drift term is shown to be negligible and the isotropic
hot-electron Fermi distribution is established under the 
  scattering during the linearly polarized laser pulse investigated here.
However, in the case with low pump-phonon energy, the drift term is important
and leads to a net momentum transfer from the electric field 
to electrons. Due to this net momentum and the dominant Coulomb scattering, a
drifted Fermi distribution different from the one established under static
electric field is found to be established in several hundred femtoseconds. We
also show that the Auger process investigated in the literature involving only
the diagonal terms of density matrices is forbidden by the dynamic 
screening. However, we propose an Auger process involving the interband
coherence  and show that it contributes to the dynamics of carriers when the
pump-photon energy is low. In addition, the anisotropically momentum-resolved hot-phonon
temperatures due to the linearly polarized light are also investigated, with the
underlying physics revealed.  
\end{abstract}


\maketitle

\section{INTRODUCTION}
Graphene is a promising material which has received intensive attention both
theoretically and experimentally due to its excellent transport and optical
properties.\cite{NetoReview,DSarmaReview,Peres,Schwierz,Young,AvourisNanoLett,AbergelAdvPhy,ChoiCRSSMS,Kotov,Castro,Bonaccorso,Avouris} 
Especially, its linear band structure and zero band gap lead to unique optical and
electrical properties, making it suitable for various optoelectronic
applications.\cite{Bonaccorso,Avouris} A comprehensive understanding of the
optical and electrical properties is one of the prerequisites for its potential
applications.

Experimentally, the time-resolved pump-probe measurement is widely used to
investigate the dynamics of carriers in 
graphene.\cite{Dawlaty,Wang,Huang,BreusingPRL102,ShangACSNano,GeorgeNanoLett,WinnerlPRL,BreusingPRB83,DSunPRL,LiPRL108,Tielrooij,ChoiApl94,Ruzicka2010,BridaArXiv,NewsonOpt,Hale,Dani2012,Plochocka}
With this method, both the differential reflection and the differential
transmission (DT) after pump are
measured.\cite{Dawlaty,Wang,Huang,BreusingPRL102,ShangACSNano,GeorgeNanoLett,WinnerlPRL,BreusingPRB83,DSunPRL,LiPRL108,Tielrooij,ChoiApl94,Ruzicka2010,BridaArXiv,NewsonOpt,Hale,Dani2012,Plochocka}
In the experiments with medium photoexcited electron density $N^p_e$ (around $10^{12}$~cm$^2$)
and probe-photon energy much higher than the equilibrium Fermi
energy, the positive DT is often observed with its fast relaxation of
several hundred femtoseconds followed by a slower picosecond
relaxation.\cite{Dawlaty,Wang,Hale,Ruzicka2010} The mechanism leading 
to these two distinct relaxations is attributed to the rapid carrier-phonon
thermalization and the slow decay of hot phonons,
respectively.\cite{Wang,Huang,Sun} On the other hand, in recent years,
experiments with higher photoexcited electron densities
($N^p_e>10^{13}$~cm$^{-2}$)\cite{BridaArXiv,ShangACSNano,BreusingPRB83}
show that a negative DT appears when the probe-photon energy
($>1500$~meV) is much higher than the equilibrium Fermi
energy.\cite{ShangACSNano,BreusingPRB83} Such negative DT was usually claimed
to arise from the renormalization of the single particle energy by the
electron-electron Coulomb Hartree-Fock
contribution.\cite{ShangACSNano,BreusingPRB83} It is noted that although the 
observed DT with high probe-photon energy and low photoexcited electron density 
is usually positive, negative DT was also reported with
$N^p_e=3\times10^{11}$~cm$^{-2}$ by Plochocka {\it et al.} where the sample is
70-100 layer epitaxial graphene exhibiting the Dirac-type electron
spectrum.\cite{Plochocka} In contrast to the negative DT with  
probe-photon energy much higher than equilibrium Fermi energy, other types of negative DTs
with lower probe-photon energy have already been observed in previous 
works.\cite{DSunPRL,GeorgeNanoLett} Sun {\it et al.} reported the 
  appearance of the negative DT under weak pump intensity with the probe-photon
  energy between 528 to 697 meV,\cite{DSunPRL} which is lower than twice of the
  Fermi energy ($E_F\sim350$~meV). 
This negative DT was suggested to come from the weakening of the Pauli blocking
due to the heating of the electrons by the pump pulse.\cite{DSunPRL} However,
this claim has not yet been verified by the microscopic theoretical
approach. Soon after that, the negative DT was also observed by George {\it et
  al.} in the case when the probe-photon energy is as low as a few tens of meV (with
the frequency about several terahertz).\cite{GeorgeNanoLett} With such low 
probe-photon energy, the increase of the intraband absorption is assumed to be
responsible for the negative DT.\cite{GeorgeNanoLett} In addition to these
phenomena, with the pump-probe method, the optical gain has also been reported
in optically pumped graphene very recently.\cite{LiPRL108}

In addition to these experimental works, many theoretical studies have been
devoted to get a clear comprehension on the dynamics of photoexcited carriers and
phonons.\cite{Wang,Hale,Malic,Sun,Satou,Winzer,WinzerArX1209,WinzerPRB85,Kim,TaniArxiv,ButscherAPL91,BridaArXiv}
Among these works, the simple rate equation 
approach\cite{Satou,Wang,Hale} assumes the establishment of the Fermi
distribution as a starting point and, therefore, fails to incorporate the pump process and the
buildup of the Fermi distribution. The widely used time-dependent Boltzmann
equation approach\cite{Kim,TaniArxiv,Sun} can fulfill the above aims but neglects 
the interband coherence,\cite{Malic} which should be very important during 
the pump process. More elaborative microscopic investigations with the interband
coherence included have been carried out based on many-particle density-matrix 
approach.\cite{Malic,Winzer,WinzerPRB85,WinzerArX1209} However, these theoretical
investigations are still far away from completion, as stated in the following.

Among those papers with the interband coherence
included,\cite{Malic,Winzer,WinzerPRB85,WinzerArX1209} the scattering terms 
under the Markovian  approximation with the Coulomb screening in the static
limit are applied. Nevertheless, with the static
screening, the electron-electron Coulomb scattering diverges when the 
initial and final states of the scattered electrons satisfy the condition
$\hbar v_F |{\bf q}|=|\hbar \omega|$,\cite{Sachdev,Fritz} where ${\bf q}$ and
$\hbar \omega$ are the differences in momentum and energy
between the initial and final states, respectively. To avoid this divergence, in
these papers, the delta function originated from the Markovian approximation and
reflecting the energy conservation in the Coulomb scattering term, was replaced
by a Lorentzian function 
with the broadening attributed to the non-Markovian
effect.\cite{Private} Nevertheless, under the more natural dynamic
screening,\cite{HwangPRB76,TseaAPL93,yzhou,Sun} this broadening is not
needed. This is because that the dynamic
screening also diverges in a higher order when $\hbar v_F |{\bf q}|=|\hbar
\omega|$,\cite{Muller,Wunsch} and therefore reduces the original divergent
scattering amplitudes to zero.

Another important issue frequently investigated is the influence of the Auger
process. In previous works, the interband coherence was neglected and the Auger process was
defined as the process that one of the two scattered electrons is
transferred from one band to the other while the other electron remains in the same
band.\cite{Winzer,Kim,WinzerPRB85} In the works by 
Winzer {\it et al.},\cite{Winzer,WinzerPRB85} such process was particularly
stressed. However, under the dynamic screening, this Auger process is totally forbidden because
it also satisfies the strong restriction $\hbar v_F |{\bf q}|=|\hbar \omega|$ due
to the energy and momentum conservation. However, in the case with the optical
transition, the interband coherence is generated. With the interband coherence, there are Coulomb 
scattering processes with three out of the four band indices in the Coulomb
interaction Hamiltonian being the same and the left one being different. This
is also a kind of Auger process. However, the influence of such process has not
been addressed in the literature.

Finally, in the microscopic equations obtained based on the many-particle
density-matrix framework by Knorr {\it et
  al.},\cite{Malic,WinzerPRB85,Winzer,WinzerArX1209} the pump term was obtained
under the vector potential gauge. However, the drift term which describes the
acceleration of the electrons by the electric field is not included. The drift
term may be unimportant when the pump-photon energy is very high as investigated
in these works. Nevertheless, it can 
be expected to be important when the pump-photon energy is low. On the other
hand, in the time-dependent Boltzmann 
equation approach, the drift term is included but the pump term in the equation
has to be introduced based on a Fermi-golden-rule--like approximation and hence
is suitable only for a weak pump intensity.\cite{Kim} Moreover, the Rabi 
flopping\cite{Haug} is absent in this approach. Therefore, a
general equation based on the microscopic many-particle density-matrix approach
that includes both the drift and pump terms naturally is needed.

In this paper, we perform a microscopic investigation on the dynamics of
photoexcited carriers and phonons in graphene by setting up and solving the
kinetic Bloch equations with the electron-phonon, electron-impurity and electron-electron Coulomb
scatterings explicitly included.\cite{Haug,wuReview} In our study, the dynamic
screening\cite{Hwang,Wunsch,XFWang,Ramezanali} is adopted in the Coulomb
scattering. Moreover, with the gauge invariant 
approach,\cite{Haug} both the drift and pump terms are obtained naturally.
We look into the dynamics of carriers photoexcited by the linear polarized
  laser pulses with the pump-photon energy in both near-infrared and THz regimes
under medium pump intensity. When the  
pump-photon energy is high enough, the influence of the drift term is shown to
be negligible compared to that from the pump process. Moreover, the
anisotropic photoexcited electrons tend to be isotropic 
under the scattering and an  isotropic hot-electron Fermi
distribution is established before the end of the
 pulse investigated here. On the contrary, when the pump-photon 
energy is low, the drift term causes
a net momentum transfer from the electric field to electrons. Together with the
dominant Coulomb scattering, a drifted Fermi distribution is established in
several hundred femtoseconds. Moreover, the form of the drifted Fermi distribution differs
from the one obtained under static electrics field.\cite{yzhou} Besides, we also show
that the temporal evolution of the DT measured by Hale {\it et al.} can be
well fitted with our microscopic calculation.\cite{Hale} In addition, although the
Auger process involving only the diagonal terms of the density matrices is
forbidden by the dynamic screening, we show that the Auger process involving the
interband coherence still contributes to the dynamics of carriers. However, its
contribution is important only when the pump-photon energy is
low. Similarly, it is shown that the terms neglected in the 
rotation-wave approximation,\cite{Haug}
which is widely accepted in semiconductor optics, also become important in the
case of low pump energy. 
 We also investigate the dynamics of phonons. Due to the
linearly polarized pump pulse, the ${\bf q}$-resolved hot-phonon temperatures
are anisotropic. Moreover, the anisotropic hot-phonon temperatures under high
pump energy are very different from those under low pump energy. Finally, we
investigate the negative DT by fitting the temporal evolution of DT in the
experimental work by Sun {\it et al.}.\cite{DSunPRL} Our results support their
suggestion that the negative DT comes from the weakening of the Pauli blocking
due to the heating of the electrons by the pump pulse.

This paper is organized as follows. In Sec.~{\Rmnum 2}, we set up the model,
lay out the kinetic Bloch equations and then give a simple analysis on the
kinetic Bloch equations. In Sec.~{\Rmnum 3} the results obtained numerically
from the kinetic Bloch equations are presented. We summarize in
Sec.~{\Rmnum 4}.

\section{MODEL AND FORMALISM}
We start our investigation from graphene on SiO$_2$ and SiC
substrates. In the absence of external field, the energy dispersion of graphene
near the $K$ and $K^\prime$ points can be described by the effective
Hamiltonian\cite{DiVincenzoPRB} ($\hbar\equiv 1$)
\begin{equation}
H^{\mu,{\rm AB}}_0({\bf k})=v_F(\mu\tau_xk_x+\tau_yk_y).
\label{EffectHaml}
\end{equation}
Here, $v_F$ represents the Fermi velocity, $\mu=1$ $(-1)$ for $K$ ($K^\prime$)
valley and ${\bgreek \tau}$ are the Pauli matrices in the pseudospin space
formed by the A and B sublattices of the honeycomb lattice. The eigenvalues of $H^{\mu,{\rm AB}}_0$ are
$\varepsilon_{\eta{\bf k}}=\eta v_F k$ and the corresponding eigenstates
are $\psi^{\mu}_{{\bf k}\eta}=(\mu\eta
e^{-i\mu\theta_{\bf k}},1)^{\rm T}/\sqrt{2}$ with $\eta=1$ $(-1)$ for conduction
(valence) band and $\theta_{\bf k}$ representing the polar angle of ${\bf k}$. 
The additional electron--optical-filed interaction Hamiltonian under the
vector potential gauge is given by\cite{PeresPRB} 
\begin{equation}
H^{\mu,{\rm AB}}_{\rm photon}({\bf k},t)=|e|v_F[\mu\tau_xA_x(t)+\tau_yA_y(t)]/c.
\end{equation} 
Note that this Hamiltonian is consistent with the one obtained from the
tight-binding approach.\cite{Stroucken2011,Malic2006}
Here, we only investigate the linearly polarized light with the electric field 
\begin{equation}
{\bf E}(t)=-{\bf E_0}\cos(\omega_L
t)\exp[-t^2/(2\sigma^2_t)]
\label{ElecField}
\end{equation}
along the ${\bf x}$ direction. For simplicity, the vector
potential is taken to be ${\bf A}(t)=\frac{c}{\omega_L}{\bf E_0}\sin(\omega_Lt)\exp[-t^2/(2\sigma^2_t)]$.
Our investigation is performed in the base set of the eigenstates of the
Hamiltonian [Eq.~(\ref{EffectHaml})]. Under this base set, the Hamiltonians are given by
\begin{eqnarray}
  &&\hspace{-1cm}H^\mu_0({\bf k})=v_Fk\sigma_z,\\
  &&\hspace{-1cm}H^\mu_{\rm photon}({\bf k})={|e|v_F}[{\bf A}(t)\cdot
  \hat{\bf k}\sigma_z+\mu \hat{\bf k}\times
  {\bf A}(t)\cdot \hat{\bf z} \sigma_y]/c,
  \label{PhotonElec_H}
\end{eqnarray}
with ${\bgreek \sigma}$ representing the Pauli matrices in the conduction and
valence band space, $\hat{\bf k}$ indicating the direction of ${\bf k}$ and
$\hat{\bf z}$ standing for the unit vector along the ${\bf z}$ direction.

Exploiting the nonequilibrium Green's function approach with the
gradient expansion as well as the generalized Kadanoff-Baym ansatz, the gauge
invariant kinetic equations\cite{Haug} of the carriers under the Markovian 
approximation read\cite{wuReview,yzhou,PZhang,PZhang2}
 (see Appendix~A for detailed derivation of the coherent and drift terms)
\begin{equation}
  \partial_t{{\rho}}_{\mu{\bf k}}=\left.\partial_t{{\rho}}_{\mu{\bf
        k}}\right|_{\rm coh}+\left.\partial_t{{\rho}}_{\mu{\bf k}}\right|_{\rm
    drift}+\left.\partial_t{{\rho}}_{\mu{\bf k}}\right|_{\rm scat},
\label{KEE}
\end{equation} 
in which ${\rho}_{\mu{\bf k}}$ stand for the density matrices with their
diagonal terms ${\rho}_{\mu{\bf k},\eta\eta}=f_{\mu{\bf k},\eta}$ representing
the electron distribution functions and the off-diagonal terms
${\rho}_{\mu{\bf k},1,-1}={\rho}^\ast_{\mu{\bf k},-1,1}$ standing
for the interband coherence.  The drift terms can be written as 
\begin{equation}
\left.\partial_t{{\rho}}_{\mu{\bf k}}\right|_{\rm 
  drift}=|e|{\bf E}\cdot\nabla_{\bf k}{\rho}_{\mu{\bf k}},
\end{equation}
which describe the acceleration of electrons by the laser field ${\bf E}$. The
coherent terms are given by  
\begin{equation}
  \left.\partial_t{{\rho}}_{\mu{\bf k}}\right|_{\rm
    coh}=-i[v_Fk\sigma_z+\Sigma^{\rm HF}_{\mu{\bf k}}+H^\mu_{\rm Pump},{\rho}_{\mu{\bf
      k}}].
  \label{coh}
\end{equation}
Here, $[A,B]\equiv AB-BA$ is the commutator and 
$H^\mu_{\rm Pump}=-|e|v_F\mu A_x\sin\theta_{\bf k}\sigma_y/c$
 comes from the laser field describing the pump
process. $\Sigma^{\rm HF}_{\mu{\bf k}}=-\sum_{{\bf q}}\hat{S}^{\mu}_{\bf 
  k,k-q}{\rho}_{\mu{\bf k-q}} \hat{S}^{\mu}_{\bf k-q,k}V^r({\bf q},0)$ stand for
the Hartree-Fock self-energies,\cite{ValencebandHF}
 in which $\hat{S}^{\mu}_{\bf k,k-q}$ are the
form factors with their elements being $S^\mu_{{\bf 
    k,k-q},\eta\eta^\prime}=\psi^{\mu\dagger}_{{\bf k}\eta}\psi^{\mu}_{{\bf
    k-q}\eta^\prime}$ and $V^r({\bf q},\omega)=V^0_{\bf q}/\epsilon({\bf
  q},\omega)$ represent the screened two-dimensional Coulomb potentials with $V^0_{\bf
  q}=2\pi v_F r_s /q$ being the bare Coulomb potential. Here $r_s$ stands for the
dimensionless Wigner-Seitz radius.\cite{Adam,Hwang2,Adam2,Hwang,Fratini} 
$\epsilon({\bf q},\omega)$ is the dielectric function under the dynamic
screening,\cite{Haug,Hwang2,Hwang} which is given by\cite{Hwang,Wunsch,XFWang,Ramezanali}
\begin{equation}
  \epsilon({\bf q},\omega)=1-\sum_{\mu \eta\eta^\prime {\bf k}}|S^\mu_{{\bf k},{\bf
      k+q},\eta\eta^\prime}|^2\frac{2(f_{\mu {\bf
        k},\eta}-f_{\mu {\bf k+q},\eta^\prime})V^0_{\bf
      q}}{\varepsilon_{\eta{\bf k}}-\varepsilon_{\eta^\prime{\bf
        k+q}}+\omega+i0^+}. 
\end{equation}

It is noted that the laser field ${\bf E}$ can not only pump electrons but also
accelerate them, as revealed in our gauge invariant kinetic Bloch equations
[Eq.~(\ref{KEE})]\cite{Haug} by the pump and the drift terms,
respectively. However, in previous works by Knorr {\it et
  al.},\cite{Malic,WinzerPRB85,Winzer,WinzerArX1209} the drift 
term is not included and hence the case of low pump energy, for which the drift
term is expected to be important, can not be investigated with their
  equations. In contrast, our kinetic Bloch equations can deal with both the
high and low pump energies completely.

Under the laser field with pump-photon energy $\omega_L$, the off-diagonal terms of
the density matrices are generated around the resonant absorption point (where
$2v_Fk\approx\omega_L$). They oscillate with the frequency around
$\omega_L$. Following the approach adopted in semiconductor optics,\cite{Haug} we
introduce the slowly varying interband-polarization component 
$P_{\mu{\bf k}}={\rho}_{\mu{\bf k},1,-1}e^{i\omega_L t}$. Then, the
kinetic equations can be rewritten into
\begin{eqnarray}
\partial_tf_{\mu{\bf k},\eta}&=&\eta{\rm Im}(\Omega^\ast_{\mu{\bf k}}
P_{\mu{\bf k}})+|e|{\bf E}\cdot\nabla_{\bf k}{f}_{\mu{\bf
    k},\eta}+\partial_tf_{\mu{\bf k},\eta}|_{\rm
  scat}, \nonumber \\ 
\partial_tP_{\mu{\bf k}}&=&-i{\Omega_{\mu{\bf k}}}(f_{\mu{\bf k},1}-f_{\mu{\bf
    k},-1})/2-i\nu^\mu_{\bf k}P_{\mu{\bf k}}\nonumber \\
&&\hspace{-0.06cm}\mbox{}+|e|{\bf E}\cdot\nabla_{\bf k}{P}_{\mu{\bf k}}+\partial_tP_{\mu{\bf
    k}}|_{\rm scat},
\label{BlochEquOffDia}
\end{eqnarray}
with the detuning $\nu^\mu_{\bf k}=2v_Fk-\omega_L+{\rm
  Tr}[({\Sigma^{\rm HF}_{\mu {\bf k}}+H^\mu_{\rm Pump})\sigma_z}]$ and the Rabi frequency 
\begin{eqnarray}
&&\hspace{-0.43 cm}
\Omega_{\mu{\bf k}}=-2e^{i\omega_Lt}(\Sigma^{\rm HF}_{\mu {\bf
    k},1,-1}+H^\mu_{{\rm Pump},1,-1})\nonumber \\
&&\hspace{-0.3 cm}=\mu|e|v_FE_0\sin\theta_{\bf
  k}e^{-\frac{t^2}{2\sigma^2_t}}(1-e^{2i\omega_Lt})/{\omega_L}-2\Sigma^{\rm HF}_{\mu {\bf k},1,-1}e^{i\omega_Lt}.\nonumber\\
\label{RabiLowFre}
\end{eqnarray}
When the pump-photon energy is high enough, these equations can be further
simplified by neglecting the high-frequency oscillating terms (rotation-wave
approximation)\cite{HaugKoch} and the Rabi frequency becomes  
\begin{eqnarray}
&&\hspace{-1.05cm}\Omega_{\mu{\bf k}}=\mu|e|v_FE_0\sin\theta_{\bf
    k}e^{-{t^2}/({2\sigma^2_t})}/{\omega_L}\nonumber \\
&&\hspace{-0.05cm}\mbox{}+2\sum_{{\bf q}}{S}^{\mu}_{{\bf k,k-q},1,1}{S}^{\mu}_{{\bf k-q,k},-1,-1}V^r({\bf q},0)P_{\mu{\bf k-q}}.
\end{eqnarray} 
Then, one finds that, except for the drift and scattering terms,
Eq.~(\ref{BlochEquOffDia}) is of the same form as that in semiconductor
optics.\cite{Haug} However, when the pump energy is low, the rotation-wave
approximation is no longer valid and the Rabi frequency must be calculated with
Eq.~(\ref{RabiLowFre}).

We further define a new set of matrices ${\hat{\rho}_{\mu{\bf k}}}$ with
${\hat{\rho}_{{\mu{\bf k}},\eta,\eta}}=f_{\mu{\bf k},\eta}$ and
$\hat{\rho}_{{\mu{\bf k}},1,-1}={\hat{\rho}_{{\mu{\bf k}},-1,1}}^\ast=P_{\mu {\bf k}}$.
Under such definition, $\hat{\rho}_{{\mu{\bf k}}}$ vary slowly
with time and the scattering terms in Eq.~(\ref{BlochEquOffDia}), which include the 
electron-electron Coulomb, electron-impurity, electron-optical
phonon and electron--remote-interfacial (RI) phonon scatterings, are given as
\begin{eqnarray}
&&\hspace{-0.45cm}\left.\partial_t{\hat{\rho}}_{\mu{\bf k},\eta\eta^\prime}\right|_{\rm ee}=\big\{[-\pi\hspace{-0.1cm}\sum_{{\bf
    q}\eta_1\eta_2\eta_3}\hspace{-0.1cm}S^\mu_{{\bf k},{\bf
    k-q},\eta\eta_1}\hat{\rho}_{\mu{\bf 
    k-q},\eta_1,\eta_2}^>\nonumber \\
&&\hspace{-0.1cm} \mbox{} \times S^\mu_{{\bf k-q,k},\eta_2\eta_3}\hat{\rho}_{\mu{\bf
    k},\eta_3,\eta^\prime}^<\hspace{-0.2cm}\sum_{{\bf
    k_1}\eta^\prime_1\eta^\prime_2\eta^\prime_3\eta^\prime_4,\mu^\prime}\hspace{-0.2cm}V^r({\bf
  q},\varepsilon_{\eta^\prime_1{\bf 
    k_1+q}}-\varepsilon_{\eta^\prime_4{\bf
    k_1}}) \nonumber \\
&&\hspace{-0.1cm}\mbox{}\times V^a({\bf q},\varepsilon_{\eta^\prime_2{\bf
    k_1+q}}-\varepsilon_{\eta^\prime_3{\bf k_1}})S^{\mu^\prime}_{{\bf k_1},{\bf
    k_1+q},\eta^\prime_4\eta^\prime_1}\hat{\rho}_{\mu^\prime{\bf
    k_1+q},\eta^\prime_1,\eta^\prime_2}^>\nonumber \\
&&\hspace{-0.1cm} \mbox{}\times S^{\mu^\prime}_{{\bf
    k_1+q},{\bf k_1},\eta^\prime_2\eta^\prime_3}\hat{\rho}_{\mu^\prime{\bf
    k_1},\eta^\prime_3,\eta^\prime_4}^< \nonumber \\
&&\hspace{-0.1cm}\mbox{} \times e^{i(\eta-\eta_1+\eta_2-\eta_3+\eta^\prime_4-\eta^\prime_1+\eta^\prime_2-\eta^\prime_3)\omega_L t/2}\nonumber \\
&&\hspace{-0.1cm}\mbox{}\times \delta(\varepsilon_{\eta_2{\bf k-q}}-\varepsilon_{\eta_3{\bf
    k}}+\varepsilon_{\eta^\prime_2{\bf k_1+q}}-\varepsilon_{\eta^\prime_3{\bf
    k_1}})]-[ > \longleftrightarrow < ]\big\}\nonumber\\
&&\hspace{-0.1cm}\mbox{}+\big\{\eta
  \longleftrightarrow \eta^\prime\big\}^\ast,\label{CoulombScatteringTerm}\\
&&\hspace{-0.45cm}\left.\partial_t{\hat{\rho}}_{\mu{\bf k},\eta\eta^\prime}\right|_{\rm ei}=\big\{[-\pi\sum_{{\bf
    q}\eta_1\eta_2}S^\mu_{{\bf k},{\bf k-q},\eta\eta_1}\hat{\rho}_{\mu{\bf
    k-q},\eta_1,\eta_2}^>\nonumber \\ && \mbox{} \times S^\mu_{{\bf k-q,k},\eta_2\eta_2}\hat{\rho}_{\mu{\bf
    k},\eta_2,\eta^\prime}^<|U(\varepsilon_{\eta_2{\bf
    k}}-\varepsilon_{\eta_2{\bf k-q}})|^2 e^{i({\eta-\eta_1})\omega_L t/{2}} \nonumber \\
&&\hspace{-0.1cm}\mbox{}\times \delta(\varepsilon_{\eta_2{\bf k-q}}-\varepsilon_{\eta_2{\bf
    k}})]-[ > \longleftrightarrow < ]\big\}+\big\{\eta
  \longleftrightarrow \eta^\prime\big\}^\ast,\label{eiscattering}\\
&&\hspace{-0.45cm}\left.\partial_t{\hat{\rho}}_{\mu{\bf k},\eta\eta^\prime}\right|_{\rm ep}=\big\{-\pi\sum_{{\bf
      q}\eta_1\eta_2\eta_3\mu^\prime\lambda\pm}M^{\mu\mu^\prime\lambda}_{{\bf k},{\bf
      k-q},\eta\eta_1}M^{\mu^\prime\mu\lambda}_{{\bf
      k-q,k},\eta_2\eta_3}\nonumber \\ 
  &&\hspace{-0.1cm}\mbox{}\times(N^{\pm\lambda}_{{\bf q}}\hat{\rho}_{\mu^\prime{\bf k-q},\eta_1,\eta_2}^>\hat{\rho}_{\mu{\bf
      k},\eta_3,\eta^\prime}^<-N^{\mp\lambda}_{-{\bf q}}\hat{\rho}_{\mu^\prime{\bf k-q},\eta_1,\eta_2}^<\hat{\rho}_{\mu{\bf
      k},\eta_3,\eta^\prime}^>)\nonumber \\
&& \hspace{-0.1cm}\mbox{}\times e^{i({\eta-\eta_1+\eta_2-\eta_3})\omega_L
    t/{2}}\delta(\varepsilon_{\eta_2{\bf k-q}}-\varepsilon_{\eta_3{\bf
      k}}\pm\omega_{q\lambda})\big\}\nonumber \\ && \mbox{}+\big\{\eta \longleftrightarrow
  \eta^\prime\big\}^\ast,\label{epscattering}
\end{eqnarray}
in which $[ > \longleftrightarrow < ]$ stands for the same term in the previous
$[\mbox{ }]$ but interchanging $>\longleftrightarrow<$, and $\{\eta \longleftrightarrow
\eta^\prime\}^\ast$ represents the conjugate of the same term in the previous $\{\mbox{ }\}$ except
the exchange of $\eta$ and $\eta^\prime$. In
Eq.~(\ref{epscattering}), $N^{\pm\lambda}_{\bf
  q}=n^{\lambda}_{\pm{\bf q}}+{1}/{2}\pm{1}/{2}$ with
$n^{\lambda}_{\bf q}$ standing for the distribution function of the phonon in
branch $\lambda$; $V^a({\bf q},\varepsilon_{\eta^\prime{\bf k+q}}-\varepsilon_{\eta{\bf
    k}})=[V^r({\bf q},\varepsilon_{\eta^\prime{\bf k+q}}-\varepsilon_{\eta{\bf
    k}})]^\ast$; $\hat{\rho}_{\mu{\bf k}}^>=1-\hat{\rho}_{\mu{\bf k}}$ and
$\hat{\rho}_{\mu{\bf k}}^< =\hat{\rho}_{\mu{\bf k}}$. The optical phonons
include the transverse optical phonons (KO) near the
${\rm K}$ $({\rm K}^\prime)$ point and the longitudinal (LO) as well as
transverse optical (TO) phonons near the ${\rm \Gamma}$ point. The detailed
forms of the scattering matrix elements $U(\varepsilon_{\eta_2{\bf
    k}}-\varepsilon_{\eta_2{\bf k-q}})$ and $M^{\mu\mu^\prime\lambda}_{{\bf
    k},{\bf k-q},\eta\eta_1}$ are given in Appendix~B.

In this work, the ${\bf q}$-resolved dynamics of the RI and optical phonons are
also studied, while the acoustic phonons are set to be in equilibrium with the
environment at temperature $T_0$. The kinetic
equations of the hot phonons are 
\begin{equation}
\partial_t n^{\lambda}_{\bf q} = \partial_t n^\lambda_{\bf q}|_{\rm ep} + \partial_t n^\lambda_{\bf q}|_{\rm pp},
\label{hotphonon}
\end{equation}
in which $\partial_t n^\lambda_{\bf q}|_{\rm ep}$ come from the
carrier-phonon scattering and $\partial_t n^\lambda_{\bf q}|_{\rm pp}$
 describe the anharmonic decay of hot phonons in relaxation time
 approximation. They are given by
\begin{eqnarray}
\left.\partial_t n^\lambda_{\bf q}\right|_{\rm ep}=&& {\rm Re}\big[2\pi\sum_{{\bf
    k}\eta_1\eta_2\eta_3\eta_4\mu\mu^\prime}M^{\mu\mu^\prime}_{{\bf k+q},{\bf
    k},\eta_4\eta_1}M^{\mu^\prime\mu}_{{\bf k,k+q},\eta_2\eta_3}\nonumber \\
&&\hspace{-1.8 cm}\mbox{}\times (\hat{\rho}_{\mu^\prime{\bf k},\eta_1,\eta_2}^>\hat{\rho}_{\mu{\bf
    k+q},\eta_3,\eta_4}^<N^{+\lambda}_{\bf q}-\hat{\rho}_{\mu^\prime{\bf k},\eta_1,\eta_2}^<\hat{\rho}_{\mu{\bf
    k+q},\eta_3,\eta_4}^>N^{-\lambda}_{-{\bf q}})\nonumber \\
&&\hspace{-1.8 cm}\mbox{}\times e^{i({\eta_2-\eta_1+\eta_4-\eta_3})\omega_L t/2}\delta(\varepsilon_{\eta_3{\bf k+q}}-\varepsilon_{\eta_2{\bf k}}-\omega_{q\lambda})\big], \label{nph}\\
\left.\partial_t n^\lambda_{\bf q}\right|_{\rm pp}=&&-({n^\lambda_{\bf
    q}-n^{0}_{\lambda\bf q}})/{\tau_{\rm pp}},
\end{eqnarray}
where $\tau_{\rm pp}$ is the phenomenological relaxation time
from the phonon-phonon scattering and $n^0_{\lambda,{\bf q}}$ is the number of the 
$\lambda$ branch phonons at environmental temperature $T_0$.

Before  we show our numerical results, we first give a simple analysis on the
Auger process.  If only the diagonal terms of the density matrices (i.e.,
$\eta=\eta^\prime$, $\eta_1=\eta_2$, $\eta_3=\eta^\prime$, $\eta^\prime_1=\eta^\prime_2$ and
 $\eta^\prime_3=\eta^\prime_4$) are involved, the Coulomb scattering terms
 become 
\begin{eqnarray}
&&\hspace{-0.45cm}\left.\partial_t{f}_{\mu{\bf k},\eta}\right|_{\rm ee}=[-2\pi\hspace{-0.1cm}\sum_{{\bf
    q}\eta_1}\hspace{-0.1cm}S^\mu_{{\bf k},{\bf
    k-q},\eta\eta_1}f_{\mu{\bf 
    k-q},\eta_1}^>\nonumber \\
&&\hspace{-0.2cm} \mbox{} \times S^\mu_{{\bf k-q,k},\eta_1\eta}f_{\mu{\bf
    k},\eta}^<\hspace{-0.2cm}\sum_{{\bf
    k_1}\eta^\prime_1\eta^\prime_2,\mu^\prime}\hspace{-0.2cm}V^r({\bf
  q},\varepsilon_{\eta^\prime_1{\bf 
    k_1+q}}-\varepsilon_{\eta^\prime_2{\bf
    k_1}}) \nonumber \\
&&\hspace{-0.2cm}\mbox{}\times V^a({\bf q},\varepsilon_{\eta^\prime_1{\bf
    k_1+q}}-\varepsilon_{\eta^\prime_2{\bf k_1}})S^{\mu^\prime}_{{\bf k_1},{\bf
    k_1+q},\eta^\prime_2\eta^\prime_1}f_{\mu^\prime{\bf
    k_1+q},\eta^\prime_1}^>f_{\mu^\prime{\bf
    k_1},\eta^\prime_2}^<\nonumber \\
&&\hspace{-0.2cm} \mbox{}\times S^{\mu^\prime}_{{\bf
    k_1+q},{\bf k_1},\eta^\prime_1\eta^\prime_2}\delta(\varepsilon_{\eta_1{\bf k-q}}-\varepsilon_{\eta{\bf
    k}}+\varepsilon_{\eta^\prime_1{\bf k_1+q}}-\varepsilon_{\eta^\prime_2{\bf
    k_1}})]\nonumber \\
&&\hspace{-0.2cm}\mbox{}-[>\longleftrightarrow<].\label{CoulombScatteringTermDia}
\end{eqnarray}
From this equation, one finds that it describes the scattering process with one
electron being scattered between
bands $\eta$ and $\eta_1$ while the other electron between bands
$\eta^\prime_2$ and $\eta^\prime_1$ under the energy conservation condition
$\delta(\varepsilon_{\eta_1{\bf 
    k-q}}-\varepsilon_{\eta{\bf k}}+\varepsilon_{\eta^\prime_1{\bf
    k_1+q}}-\varepsilon_{\eta^\prime_2{\bf k_1}})$. Moreover, the vertices of
the involved two scattering processes, 
$V^r({\bf q},\varepsilon_{\eta^\prime_1{\bf
    k_1+q}}-\varepsilon_{\eta^\prime_2{\bf k_1}})S^{\mu^\prime}_{{\bf k_1},{\bf
    k_1+q},\eta^\prime_2\eta^\prime_1}S^\mu_{{\bf k,k-q},\eta\eta_1}$ and
$V^a({\bf q},\varepsilon_{\eta^\prime_1{\bf
    k_1+q}}-\varepsilon_{\eta^\prime_2{\bf k_1}})S^\mu_{{\bf
    k-q,k},\eta_1\eta}S^{\mu^\prime}_{{\bf k_1+q},{\bf
    k_1},\eta^\prime_1\eta^\prime_2}$, are conjugated with each other.  
For the Auger-type scattering process here, it means that three out of the four
involved band indices ($\eta$, $\eta_1$, $\eta^\prime_1$ and $\eta^\prime_2$)
are the same and the left one is different, corresponding to the case that one
of the two scattered electrons is transferred from one band to the other while
the other electron remains in the same band. Then, taking 
$\eta_1=\eta^\prime_1=\eta^\prime_2\neq\eta$ for example and substituting 
the energy dispersion, the requirement of energy conservation
becomes $v_F(|{\bf k-q}|+|{\bf k}|+|{\bf k_1+q}|-|{\bf k_1}|)=0$. Since $|{\bf
  k-q}|+|{\bf k}|\geq|{\bf k-q-k}|=q=|{\bf k_1+q-k_1}|\geq|{\bf k_1}|-|{\bf
  k_1+q}|$, one finds that the condition 
\begin{equation}
  v_Fq=|\varepsilon_{\eta^\prime_1{\bf k_1+q}}-\varepsilon_{\eta^\prime_2{\bf
      k_1}}|
\label{QOmegaRelation}
\end{equation}
must be satisfied. Although this condition is obtained in the special case
$\eta_1=\eta^\prime_1=\eta^\prime_2\neq\eta$, one can prove that it is valid as long as
the band indices in the delta function is Auger type. On the other hand, it has
been shown that, under this condition, the dielectric function under the dynamic
screening $\epsilon({\bf q},\varepsilon_{\eta^\prime_1{\bf
    k_1+q}}-\varepsilon_{\eta^\prime_2{\bf k_1}})$ also
diverges.\cite{Wunsch,Muller} Therefore, both $V^r({\bf  
  q},\varepsilon_{\eta^\prime_1{\bf k_1+q}}-\varepsilon_{\eta^\prime_2{\bf
    k_1}})$ and $V^a({\bf q},\varepsilon_{\eta^\prime_1{\bf
    k_1+q}}-\varepsilon_{\eta^\prime_2{\bf k_1}})$ are reduced to zero and
the corresponding processes are forbidden. Nevertheless, if the scatterings involving the
off-diagonal terms are taken into consideration, terms like
\begin{eqnarray}
  &&\hspace{-0.45cm}-\pi\hspace{-0.1cm}\sum_{{\bf q}\eta_1\eta_2\eta_3}\sum_{{\bf
      k_1}\eta^\prime_1\eta^\prime_2\bar{\eta}^\prime_1\bar{\eta}^\prime_2\mu^\prime}\hspace{-0.1cm}S^\mu_{{\bf k},{\bf k-q},\eta\eta_1}\hat{\rho}_{\mu{\bf 
      k-q},\eta_1,\eta_2}^>\nonumber \\
  &&\hspace{-0.1cm} \mbox{} \times S^\mu_{{\bf k-q,k},\eta_2\eta_3}\hat{\rho}_{\mu{\bf
      k},\eta_3,\eta^\prime}^<V^r({\bf
    q},\varepsilon_{\eta^\prime_1{\bf
      k_1+q}}-\varepsilon_{\bar{\eta}^\prime_1{\bf k_1}}) \nonumber \\ 
  &&\hspace{-0.1cm}\mbox{}\times V^a({\bf q},\varepsilon_{\eta^\prime_2{\bf
      k_1+q}}-\varepsilon_{\bar{\eta}^\prime_2{\bf k_1}})S^{\mu^\prime}_{{\bf k_1},{\bf
      k_1+q},\bar{\eta}^\prime_1\eta^\prime_1}\hat{\rho}_{\mu^\prime{\bf
      k_1+q},\eta^\prime_1,\eta^\prime_2}^>\nonumber \\
  &&\hspace{-0.1cm} \mbox{}\times S^{\mu^\prime}_{{\bf
      k_1+q},{\bf k_1},\eta^\prime_2\bar{\eta}^\prime_2}\hat{\rho}_{\mu^\prime{\bf
      k_1},\bar{\eta}^\prime_2,\bar{\eta}^\prime_1}^<
  e^{i({\eta-\eta_1+\bar{\eta}^\prime_1-\eta^\prime_1})\omega_L
    t/{2}}\delta_{\bar{\eta}^\prime_1,-\eta^\prime_1\eta\eta_1}\nonumber \\ 
  &&\hspace{-0.1cm}\mbox{}\times \delta(\varepsilon_{\eta_2{\bf k-q}}-\varepsilon_{\eta_3{\bf
      k}}+\varepsilon_{\eta^\prime_2{\bf k_1+q}}-\varepsilon_{\bar{\eta}^\prime_2{\bf
      k_1}})\delta_{\bar{\eta}^\prime_2,\eta^\prime_2\eta_2\eta_3}\label{CoulombScatteringTermOffDia}
\end{eqnarray}
also exist in $\left.\partial_t{\hat{\rho}}_{\mu{\bf
      k},\eta\eta^\prime}\right|_{\rm ee}$. Here,
$\delta_{\bar{\eta}^\prime_1,-\eta^\prime_1\eta\eta_1}$ indicates an Auger type  
scattering process for the scattering with the vertex $V^r({\bf q},\varepsilon_{\eta^\prime_1{\bf k_1+q}}-\varepsilon_{\bar{\eta}^\prime_1{\bf k_1}})S^\mu_{{\bf k},{\bf
      k-q},\eta\eta_1}S^{\mu^\prime}_{{\bf k_1},{\bf
      k_1+q},\bar{\eta}^\prime_1\eta^\prime_1}$. However,
$\delta_{\bar{\eta}^\prime_2,\eta^\prime_2\eta_2\eta_3}$ guarantees that the
scattering process restricted by the energy conversion [whose
vertex is $V^a({\bf q},\varepsilon_{\eta^\prime_2{\bf
      k_1+q}}-\varepsilon_{\bar{\eta}^\prime_2{\bf k_1}})S^{\mu^\prime}_{{\bf
      k_1+q},{\bf k_1},\eta^\prime_2\bar{\eta}^\prime_2}S^\mu_{{\bf
      k-q,k},\eta_2\eta_3}$] is non-Auger type, which means that the requirement
  Eq.~(\ref{QOmegaRelation}) is no longer necessary. Therefore, these
  scattering terms are permitted and are expected to contribute to the dynamic
  of carriers. Moreover, for the corresponding scattering term
  [Eq.~(\ref{CoulombScatteringTermOffDia})], there is a prefactor 
$\exp[{i(\eta-\eta_1+\bar{\eta}^\prime_1-\eta^\prime_1)\omega_Lt/2}]$ with
$\eta-\eta_1+\bar{\eta}^\prime_1-\eta^\prime_1\neq0$. Since the 
other factors in the scattering term all vary slowly with time, the
contribution of such scattering term oscillates very fast when
$\omega_L$ is large and hence it has little effect on the dynamics of
carriers. However, when $\omega_L$ is small enough,  
this kind of Auger process is expected to have marked influence on the
evolution of electron distribution function.

\section{RESULTS}
The kinetic Bloch equations, Eqs.~(\ref{KEE}) and (\ref{hotphonon}) with all terms
included are numerically solved following the method detailed in
Ref.~\onlinecite{Weng}, except the drift 
term. In this paper, the drift term is treated up to the third order to suppress the
noise (see Appendix~C). Then, the temporal
evolutions of the carrier and phonon distribution functions are
obtained. From Eq.~(\ref{BlochEquOffDia}), one finds
 that $f_{\mu{\bf k},\eta}=f_{-\mu{\bf
      k},\eta}$ and $P_{\mu{\bf k}}=-P_{-\mu{\bf k}}$.  Therefore, we only show the results
in $K$ valley ($\mu=1$) with the valley index $\mu$ no longer appearing for
simplicity. With the evolution of carrier distribution, the 
temporal evolution of the optical transmission at the 
probe-photon energy $\omega_{\rm pr}$ is calculated
from\cite{Dawlaty,Rana,Dawlaty2} 
\begin{equation}
T_{\omega_{\rm pr}}(t)=|1+N_{\rm 
  lay}\sigma_{\omega_{\rm pr}}(t)\sqrt{\mu_0/\epsilon_0}/(1+n_{\rm ref})|^{-2},
\label{Trans}
\end{equation}
where $n_{\rm ref}$ is the refractive index of the substrate and $N_{\rm lay}$
is the number of graphene layers. Here, the corresponding interband optical
conductivity, which is determined by carrier distributions, is given
by\cite{Dawlaty,PeresPRB,Scharf,anisotropic}
\begin{equation}
\sigma_{\omega_{\rm pr}}(t)=-\frac{e^2}{4\pi}\int^{2\pi}_0d\theta_{\bf
  k_\omega}\sin^2(\theta_{\bf k_\omega})[{f}_{{\bf k_\omega},1}(t)-{f}_{{\bf k_\omega},-1}(t)],
\label{conductivity}
\end{equation} 
in which $|{\bf k_\omega}|=\omega_{\rm pr}/(2v_F)$ is the resonant absorption
point. DT is calculated from $\Delta T_{\omega_{\rm
    pr}}(t)/T^0_{\omega_{\rm pr}}=[T_{\omega_{\rm pr}}(t)-T^0_{\omega_{\rm 
  pr}}]/T^0_{\omega_{\rm pr}}$, with $T^0_{\omega_{\rm pr}}$ representing the transmission before
the pump. We present our numerical results with material 
parameters given in Table~\ref{tab1}. The full width at half
maximum (FWHM) of the pump pulse is chosen to be 100~fs
(corresponding to $\sigma_t=42.5$~fs) and $\tau_{\rm 
  pp}=2.8$~ps.\cite{explainTaupp} The equilibrium chemical potential is taken to
be 18.6 meV (the corresponding electron density is
$1.43\times10^{11}$~cm$^{-2}$), the environmental temperature is set to
be 300~K and the impurities are absent, unless
otherwise specified.

\begin {table}[tb]
 \caption{\label{tab1} Parameters used in the computation.
}
 \begin{ruledtabular}
 \begin{tabular}{c|cc|cc}
   & $a$ & 1.42~\AA$^{\rm\,a}$ & $v_F$ & 1$\times10^8$~cm/s$^{\rm\,a}$ \\
   & $\omega_{\rm \Gamma}$ & 196~meV$^{\rm\,b}$ & $\langle D^2_{\rm
     \Gamma}\rangle$ & 45.60~eV$^2$/\AA$^2$$^{\rm\,b}$\\
   & $\omega_{K}$ & 161~meV$^{\rm\,b}$ & $\langle D^2_{K}\rangle$ &
   92.05~eV$^2$/\AA$^2$$^{\rm\,b}$ \\
   & $D$ & 19~eV$^{\rm\,c}$ & $v_{\rm ph}$ & 2$\times10^{6}$~cm/s$^{\rm\,c}$  \\
   & $\rho_m$ & 7.6$\times10^{-8}$~g/cm$^{2{\rm\,c}}$ & $d$ & 0.4~nm$^{\rm\,d,e}$\\
   \hline
 SiO$_2$  & $\omega_{\rm RI_1}$ & 59~meV$^{\rm\,e}$ & $g_1$ &
 $5.4\times10^{-3}$$^{\rm\,e}$\\
 & $\omega_{\rm RI_2}$ & 155~meV$^{\rm\,e}$ & $g_2$ & $3.5\times10^{-2}$$^{\rm\,e}$\\
 & $r_s$ & 0.8$^{\rm\,f}$ & $n_{\rm ref}$ & 1.5$^{\rm\,g}$   \\
   \hline
 SiC  & $\omega_{\rm RI}$  & 116~meV$^{\rm\,h}$ & $g$ & $1.4\times10^{-2}$$^{\rm\,h}$ \\
               & $r_s$ & 0.4$^{\rm\,e,i}$ & $n_{\rm ref}$ & 2.6$^{\rm\,j}$ \\
 \end{tabular}\\
 \end{ruledtabular}
\vskip 0.1cm
\begin{tabular}{llll}
$^{\rm a}$ Refs.~\onlinecite{NetoReview}. &
$^{\rm b}$ Refs.~\onlinecite{Piscanec} and \onlinecite{Lazzeri}. &
$^{\rm c}$ Refs.~\onlinecite{Hwang3} and \onlinecite{Chen}. &
$^{\rm d}$ Refs.~\onlinecite{Adam2}. \\ $^{\rm e}$ Ref.~\onlinecite{Fratini}. & $^{\rm f}$
Refs.~\onlinecite{Hwang} and \onlinecite{Adam}. & $^{\rm g}$
Ref.~\onlinecite{Ghosh}.& $^{\rm h}$
Ref.~\onlinecite{Perebeinos}. \\ $^{\rm i}$
Ref.~\onlinecite{Novikov}. & $^{\rm j}$ Ref.~\onlinecite{Levinshtein}.
\end{tabular}
\end{table}

\subsection{High pump-photon  energy}
We first investigate the dynamics of carriers and phonons with
high pump-photon energy. The substrate is chosen to be SiO$_2$. In the
calculation, $\omega_L=1500$~meV (corresponding to the near-infrared light), 
unless otherwise specified.

\begin{figure}[htb]
   \includegraphics[width=8.6cm]{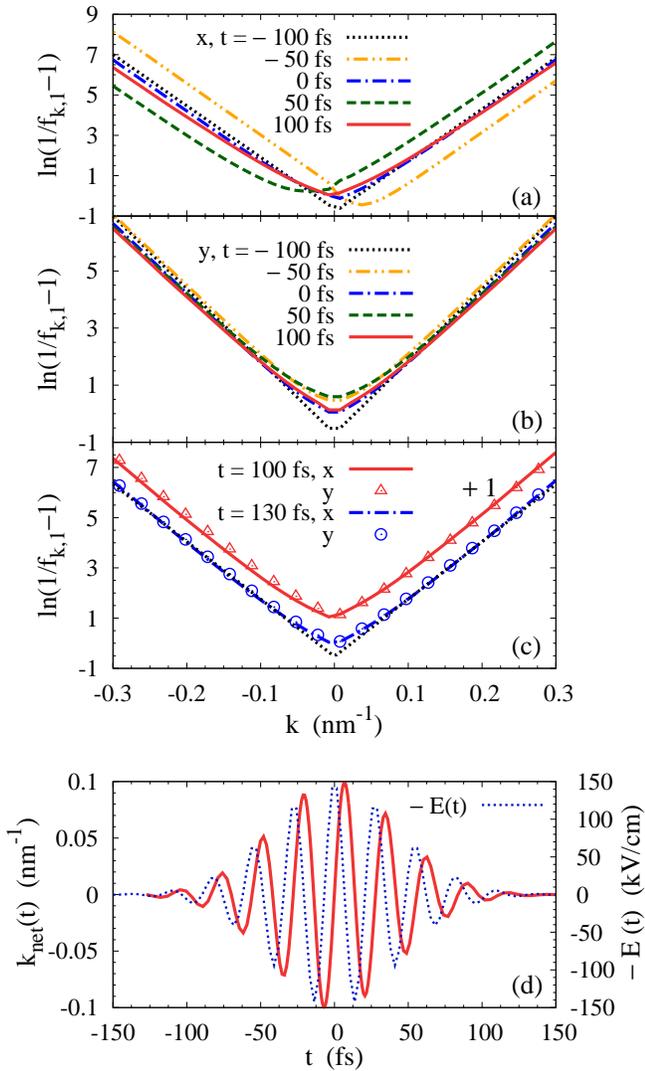}
   \caption{(Color online) (a)-(c) $\ln(f^{-1}_{{\bf k},1}-1)$ as function of $k$
     with $\omega_L=150$~meV along the ${\bf x}$ and ${\bf y}$ directions at different
     times. Here, the pump process is switched off. The black dotted curve in (c) is $(v_{\rm F}|{\bf
       k}|-\mu_1)/(k_BT_1)$ with $T_1=333$~K and $\mu_1=11.4$~meV. To
     make the results clearer, the curve for $t=100$~fs plotted in (c) is
     $\ln(f^{-1}_{{\bf k},1}-1)+1$. (d) Temporal evolution of the total
     momentum ${k}_{\rm net}(t)$ with $\omega_L=150$~meV (solid curve with the scale on the left
  hand side of the frame). The temporal 
evolution of the electric field $-E(t)$ is
     also plotted for reference (dotted curve with the scale on the right hand
     side of the frame). }
 \label{figsw1}
 \end{figure}

\subsubsection{Influence of drift term}

We first investigate the influence of the drift term on the dynamics of
electrons. To make it clearer, the pump process is switched off by
excluding $H^\mu_{\rm Pump}$ in Eq.~(\ref{coh}). In the calculation, the
maximal electric field in Eq.~(\ref{ElecField}) is taken as $E_0=150$~kV/cm and
the pump-photon energy $\omega_L=150$~meV. The $k$ dependences of
$\ln(f^{-1}_{{\bf k},1}-1)$ along the ${\bf k}_x$ and ${\bf k}_y$ directions at
different times are plotted in Figs.~\ref{figsw1}(a) and (b), respectively. The
dynamics of holes is similar and is not shown here. From the figures, one finds
that before $t=-100$~fs the influence of the electric field is
negligible. Moreover, the electron distribution at that moment is the Fermi
distribution, with 
\begin{equation}
  \ln(f^{-1}_{{\bf k},\eta}-1)=(\eta v_{\rm F}|{\bf k}|-\mu_\eta)/(k_BT_\eta)
  \label{FitFermi}
\end{equation}
as shown in the figure. Here, $T_\eta$ and $\mu_\eta$ are the electron
temperature and chemical potential in corresponding band $\eta$, respectively. 
After $t=-100$~fs, the electrons are drifted under the influence of the electric
field [Eq.~(\ref{ElecField})] along the ${\bf x}$ direction. From this
oscillating electric field, a net oscillating momentum ${\bf
  k}_{\rm net}(t)=\sum_{\bf k}{\bf k}f_{{\bf 
    k},1}(t)/\sum_{\bf k}f_{{\bf k},1}(t)$ along the ${\bf x}$
direction is gained by electrons. As a result, the
location of the minimum of $\ln(f^{-1}_{{\bf k},1}-1)$ along the ${\bf k}_x$
directions oscillates around $k=0$, as shown in Fig.~\ref{figsw1}(a). However,
the minimum along the ${\bf k}_y$ directions remains fixed at $k=0$ 
but its magnitude oscillates, as revealed in Fig.~\ref{figsw1}(b). To further illustrate the
influence of the electric field on the net momentum ${\bf
  k}_{\rm net}(t)$, we plot the temporal evolution of 
$k_{\rm net}$ in Fig.~\ref{figsw1}(d). 
The electric field $-E(t)$ is also plotted in the same figure
 for reference. One
finds that its phase is always $\pi/2$ in advance of $k_{\rm net}$.
 Moreover, after the pulse ($t>100$~fs, when $|E|<0.1|E_0|$), the net momentum
gained from the electric field tends to zero. In addition, it is found that for
curves at $t=-100$, $0$ and $100$~fs in Fig.~\ref{figsw1}(a), the magnitudes of
slopes of the curves away from the minima decrease with the temporal evolution,
indicating that electrons are heated by the drift pulse [see
Eq.~(\ref{FitFermi})]. Then, with the negligible net momentum, the isotropic
hot-electron Fermi distribution is expected to be established under the Coulomb
scattering after the pulse. To show this, we plot the $k$ 
dependences of $\ln(f^{-1}_{{\bf k},1}-1)$ along the ${\bf k}_x$ and ${\bf
  k}_y$ directions at $t=100$ and $130$~fs in Fig.~\ref{figsw1}(c). One finds
that at $t=100$~fs, the curve is almost linear with $k$ for each direction,
which indicates the establishment of the Fermi distribution 
along that direction. To further show the buildup of the
  isotropic Fermi distribution, one fits the distribution with
  Eq.~(\ref{FitFermi}) along the direction $\theta_{\bf k}$ and obtains the
  corresponding hot-electron temperature $T_{\eta\theta_{\bf k}}$ and chemical potential
$\mu_{\eta\theta_{\bf k}}$. Then, the establishment of the isotropic Fermi
distribution is identified when $T_{\eta\theta_{\bf k}}$ along different
directions are very close. Here, it is found that at $t=130$~fs the relative
difference of the temperature increase $T_{1\theta_{\bf k}}-T_0$ along different
directions is less than 10~\%.

\begin{figure}[htb]
   \includegraphics[width=8cm]{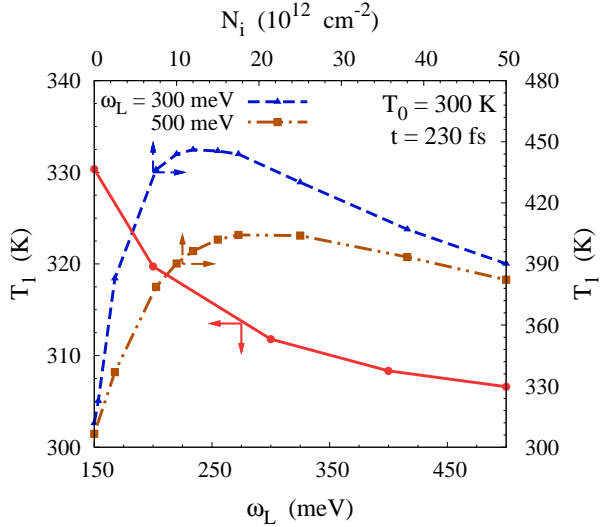}
   \caption{(Color online) Pump-photon energy $\omega_L$ and impurity density $N_i$ dependences of
     the hot-electron temperature $T_1$. Here, the hot-electron temperatures are obtained
     at $t=230$~fs and the environment temperature is $T_0=300$~K.  The impurity
     density dependence of $T_1$ are calculated with $\omega_L=300$ and
     $500$~meV and its scales are on the top and right-hand side of the frame.}
 \label{figsw2}
 \end{figure}
Then, we investigate the influences of the pump-photon energy and impurity
density on the thermalization of electrons by the drift term. We
plot the pump-photon energy and impurity density dependences of
the hot-electron temperature  $T_1$ in Fig.~\ref{figsw2}. With different
  $\omega_L$ and $N_i$, the isotropic Fermi distributions are established at
  different times. Here, $T_1$ is fitted along the ${\bf k}_x$ direction at 
$t=230$~fs, when the isotropic Fermi distribution
has just been established (relative difference between temperature increase
$T_{1\theta_{\bf k}}-T_0$ along the ${\bf k}_x$ and ${\bf k}_y$ directions
is less than 10~\%) for the slowest case in our calculation. The impurity density
dependence of $T_1$ is obtained with $\omega_L=300$ and
  $500$~meV. It is shown that $T_1$ decreases monotonically with the
increase of the pump-photon energy. Moreover, a peak appears in the impurity
density dependence as shown in Fig.~\ref{figsw2}. These phenomena can be well
understood from the Boltzmann equation under the relaxation time approximation:
$\partial_tf_{{\bf k},\eta}=|e|{\bf E}\cdot \nabla_{\bf k}{f}_{{\bf
    k},\eta}-(f_{{\bf k},\eta}-f^0_{{\bf k},\eta})/\tau_p$, with $f^0_{{\bf
    k},\eta}$ being the equilibrium distribution and $\tau_p$ standing for the
relaxation time. From this equation, one obtains the conductivity under the AC
electric field ${\bf E}=E_0\exp{(-i\omega_Lt)}\hat{\bf x}$ as 
\begin{eqnarray}
  \hspace{-0.05cm}\sigma(\omega_L)=\frac{e^2v^2_F}{\pi}\int^\infty_0 dk
  k\frac{\tau_p}{1+(\omega_L\tau_p)^2}\sum_\eta(-\frac{\partial f_{{\bf
        k},\eta}}{\partial \varepsilon_{\eta {\bf k}}}), 
\label{intraSimga}
\end{eqnarray}
which is the same as the intraband conductivity given in the 
literature.\cite{Satou,Dawlaty2,PeresPRB} From this equation, it is found that the
conductivity and hence the energy absorbed from the electric field decreases
monotonically with the increase of $\omega_L$ and a peak exists in its $N_i$ (or
equivalently $1/\tau_p$) dependence. These phenomena can 
also be understood more physically. One finds from Eq.~(\ref{intraSimga}) that 
the conductivity tends to zero in the limit of no scattering or infinitely
large scattering. This can be comprehended as the effect of the electric field on the
electron is totally cancelled out in one oscillation period in the case without
scattering and the electrons can not be drifted by the electric field in the
limit of infinitely large scattering strength. Then, in the case with
finite scattering strength, a peak must appear as shown in Fig.~\ref{figsw2}. 
As for the decrease of $T_1$ with the increasing
$\omega_L$, it is understood as the electron becomes more and more difficult to
follow the laser field with the increase of the oscillation frequency.

To further compare the influence from the drift and the pump terms, we
also perform the calculation with the pump process included but the drift term excluded. In
the calculation, $\omega_L=500$~meV, $N_i=0$ and the other parameters remain the
same. At the time in Fig.~\ref{figsw2}, i.e., $t=230$~fs, the isotropic Fermi
distribution is checked to have already 
been established and the obtained hot-electron temperature is $T_1=854$~K,
much larger than 7~K established by the drift term only. In the sample
usually reported in the experiments,\cite{Pi2010,Dorgan2010}
the mobility is larger than 
$1000$~cm$^2$/Vs ($N_i<2.5\times10^{12}$~cm$^{-2}$). Within this impurity
density range, the temperature increase due to the drift term is less than
$37$~K. This is much smaller than that from the pump process (the influence of
impurity on the pump process is small with such pump intensity and impurity
density range). Moreover, since the thermalization due to the drift effect 
decreases with increasing $\omega_L$, the drift term is negligible in the
case with pump-photon energy higher than 500~meV. This is consistent with
the investigation in semiconductors where the energy of the pump photon is very
high due to the large band gap.\cite{Haug} However, in the case with very
  low pump-photon energy as we will investigate in the next subsection, the drift term is
  expected to be very important, especially for the case where the electric
  field does not change direction during the pulse and a large net momentum is
  transferred to the electrons via the drift term.

\subsubsection{Dynamics of electrons} 
\begin{figure}[tbp]
  \includegraphics[width=8cm]{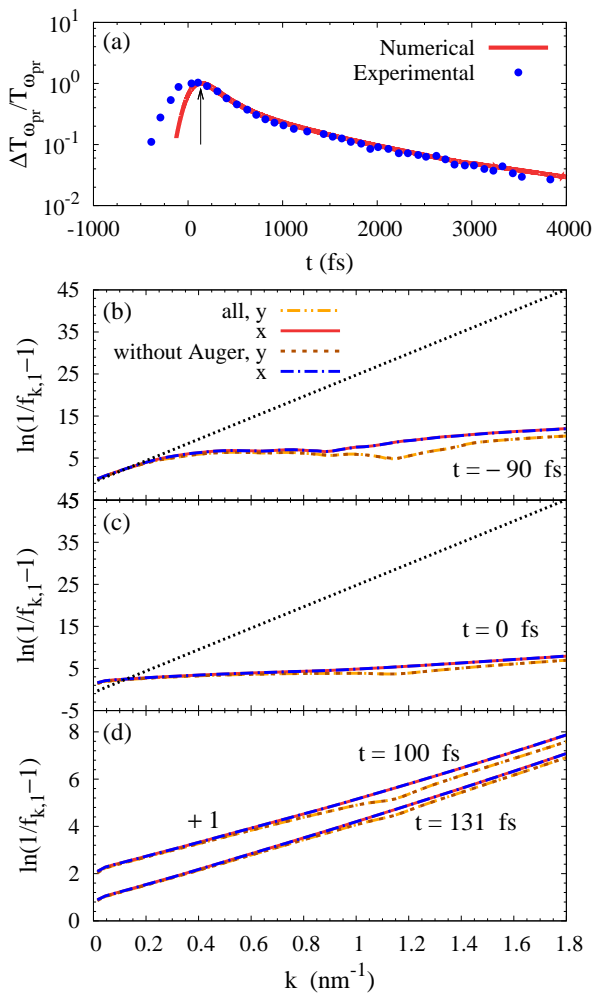}
  \caption{(Color online) (a) DT from the numerical
  results compared with the experimental data in single-layer
  graphene.\cite{Hale} The results are normalized 
  as  Hale {\it et al.}.\cite{Hale} The arrow in the figure indicates $t=133$~fs, which is the
  beginning of the relaxation. (b)-(d) $\ln(f^{-1}_{{\bf
      k},1}-1)$ as function of $k$ along the ${\bf x}$ and ${\bf y}$ directions
  at $t=-90$, 0, 100 and 131~fs. The black dotted curves in (b) and (c) are the
  case before the pump pulse. The curve for $t=100$~fs plotted in (d) is
  $\ln(f^{-1}_{{\bf k},1}-1)+1$.  The cases without the Auger process are also plotted in (b)-(d).} 
\label{figsw3}
\end{figure}

Before the investigation on electron dynamics, we first calculate the DT and
compare it with the experimental data in single-layer graphene by Hale {\it et al.}\cite{Hale}
[Fig.~\ref{figsw3}(a)]. It is noted that this experiment has been fitted in our
previous work.\cite{Sun} However, in that paper, the Fermi 
distribution was assumed to be established at all the time and the hot phonons are averaged with 
a fitting parameter $E_{\rm max}$, which describes the upper energy limit of the hot
carriers being able to emit phonons. In this paper, with the inclusion of 
the pump term and the ${\bf q}$-resolved hot phonons, the buildup of the Fermi
distribution can be studied elaborately and the fitting parameter $E_{\rm max}$ is no longer
needed. Then, the influence of the approximations in our previous
paper can be checked. In the calculation here, the probe-photon energy
$\omega_{\rm pr}=1100$~meV, the FWHM is taken to be 180~fs and the pump-photon energy
$\omega_L=1500$~meV, as indicated in the experiment. It is noted that with
$N_i=0$ and such high pump-photon
energy, the drift effect of the laser field is negligible as shown
in the previous subsection. In addition to these parameters, the fitting
parameter $\tau_{\rm pp}=2.8$~ps and the maximal electric field $E_0=300$~kV/cm
which leads to the maximal electron density being $2.7\times 10^{12}$~cm$^{-2}$.
This electron density is comparable with the experimental estimation and that in
our previous paper ($4.6\times10^{12}$~cm$^{-2}$). For the relaxation of DT which begins from
$t=133$~fs, as shown in Fig.~\ref{figsw3}(a), one finds that our numerical
results agree very well with the experimental data with a fast relaxation
followed by a slow one with the relaxation times being 0.28 and 1.33~ps,
respectively. The discrepancy between our numerical results with the
experimental data at the initial increase of the DT is understood to come from the
approximation in our model where the interference between the pump and probe pulses
is neglected and the Markovian approximation\cite{Haug} is applied.

Then, we investigate the dynamics of electrons, especially the
buildup of the Fermi distribution under the linearly polarized light. We plot 
$\ln(f^{-1}_{{\bf k},1}-1)$ as function of $k$ along different 
directions at different times in Figs.~\ref{figsw3}(b)-(d). With the strong
scattering and the negligible drift effect, the distribution 
is center symmetric and hence the distribution functions along the $-{\bf x}$ and 
$-{\bf y}$ directions are not shown. Besides, the dynamics of holes is 
almost the same\cite{Sun} and is not plotted here. From 
Fig.~\ref{figsw3}(b), one finds that the distribution of the pumped electrons
are anisotropic and mainly around $k=1.14$~nm$^{-1}\approx
\omega_L/(2\hbar v_F)$ along the ${\bf y}$-axis as shown by the valley there. This
comes from the linearly polarized pump light as shown by $\sin{\theta_{\bf k}}$ in
Eq.~(\ref{RabiLowFre}). During the pump process, the 
scattering also plays an important role by smearing out the valleys and making
the distribution tend to be isotropic as shown in
Figs.~\ref{figsw3}(b) and (c). Moreover, it is found that the slopes of the curves around
the Dirac point decrease with the temporal 
evolution, indicating their thermalizations under the scattering. At $t=131$~fs
shown in Fig.~\ref{figsw3}(d), one  
finds that the curves become almost linear with $k$ for each direction. By fitting the curves with
Eq.~(\ref{FitFermi}), the obtained hot-electron temperature $T_{1\theta_{\bf k}}$ along the
${\bf x}$ and ${\bf y}$ directions are 2160 and 2221 K, respectively with the
relative difference for these two directions being less than
3~\%. By noticing that the pulse [Eq.~(\ref{ElecField})] ends after $t=165$~fs
(when $|E|<0.1|E_0|$), one concludes that in the case
with such long pump pulse width, the hot-electron Fermi distribution is
established during the pulse and once it is established, it is an isotropic one.
Considering that the relaxation of the DT begins at 133~fs which is  after the
establishment of the Fermi distribution, the approximation of the buildup of the
Fermi distribution in our previous work is acceptable. However, this
approximation leads to the inaccuracy of the hot-electron temperature (3200~K at
the beginning of the relaxation in our previous work\cite{Sun}).

To further show the influence of the Auger process involving the interband
coherence, we also plot the results without the Auger process in Figs.~\ref{figsw3}(b)-(d). One  
finds that the exclusion of the Auger process has little influence on the
evolution of the distribution function. This is in agreement with our argument
in Sec.~{\Rmnum 2} that the Auger process which survives the dynamic screening
in the presence of the optical polarization does not contribute 
due to the high pump-photon energy. Similarly,
also due to the high pump energy, the rotation-wave approximation 
is checked to be valid (not shown in the figure).

\begin{figure*}[htb]
  \includegraphics[width=18cm]{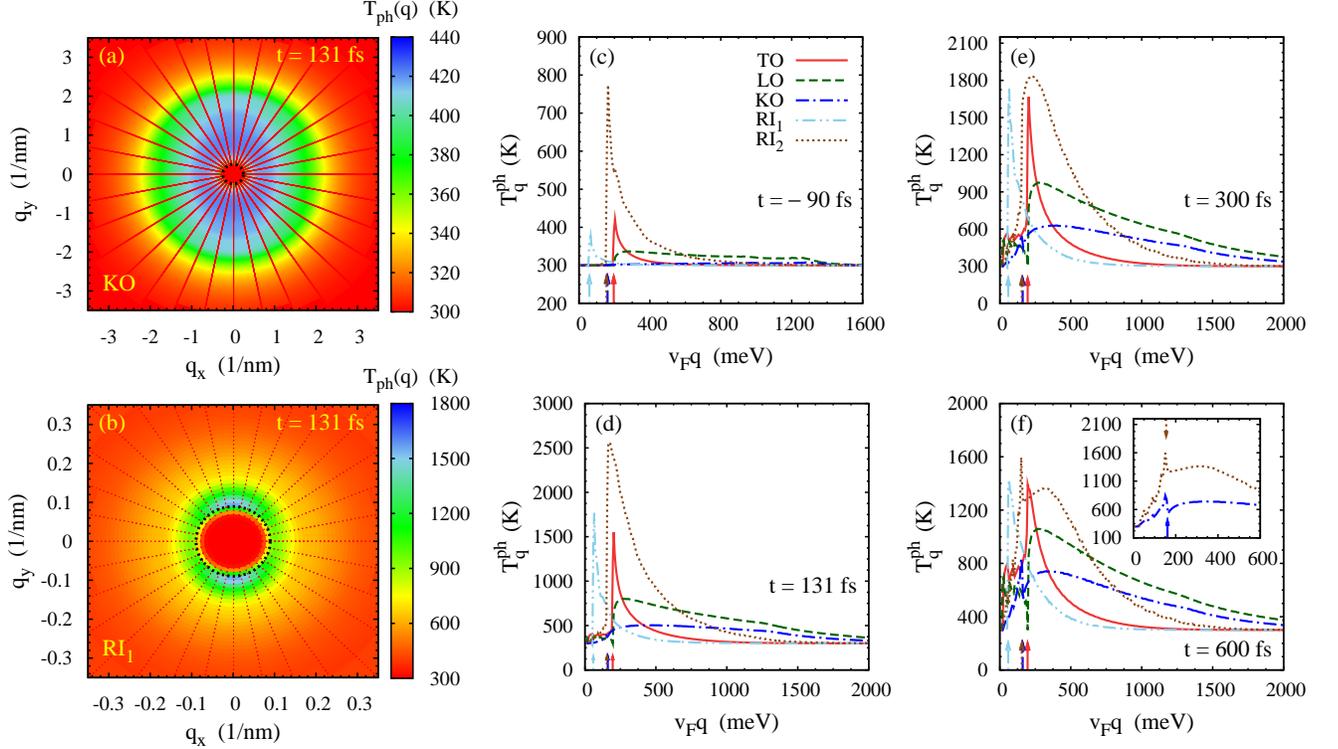}
  \caption{(Color online) (a) and (b) Typical ${\bf q}$-resolved hot-phonon
    temperatures of KO and RI$_1$ phonons at $t=131$~fs, respectively. The black
    dotted circles indicate $q=\omega_{q\lambda}/v_F$. (c)-(f) The angle
    averaged hot-phonon temperatures for different phonon branches as function of $v_Fq$ at $t=-90$, $131$, $300$
    and $600$~fs, respectively. The arrows indicate the location of the
    corresponding phonon energy $v_Fq=\omega_{q\lambda}$. It is noted that the phonon
    energies of TO and LO phonons are degenerate and we only plot the arrow for
    TO phonons here. The region close to $v_F q=0$ is enlarged in the inset for
    KO and RI$_2$ phonons at $t=600$~fs. } 
\label{figsw4}
\end{figure*} 

\subsubsection{Dynamics of phonons}

With the inclusion of the ${\bf q}$-resolved hot phonons, we are able to give a more
detailed investigation on the dynamics of phonons than our previous
work.\cite{Sun} We calculate the temperature $T_{\rm ph}({\bf q})$ for phonons
with momentum ${\bf q}$ at branch $\lambda$ by $T_{\rm ph}({\bf
  q})=\omega_\lambda/[k_B\ln(1+1/n^{\lambda}_{\bf q})]$ and then, plot 
them in momentum space at $t=131$~fs for the KO and RI$_1$ phonons
in Figs.~\ref{figsw4}(a) and (b). The ${\bf q}$-resolved temperatures of the other
phonons are similar to these two phonon branches. It is shown that due to the
anisotropic distribution of electrons, 
the thermalization of phonons under the electron-phonon scattering is also
anisotropic. Moreover, it is found that the thermalization of
phonons with small $q$ ($q<\omega_{q\lambda}/v_F$, which are the phonons
within the black dotted circle) is negligible.

This can be understood from the energy conservation during the electron-phonon
scattering process [Eq.~(\ref{epscattering})], $\delta(\varepsilon_{\eta_2{\bf
    k-q}}-\varepsilon_{\eta_3{\bf k}}\pm\omega_{q\lambda})$. 
By using the relation $v_F|{\bf k-q}|-v_F|{\bf k}|\leq v_F|{\bf
  k-q-k}|=v_Fq$, one finds that the phonons with 
$q<\omega_{q\lambda}/v_F$ and $q>\omega_{q\lambda}/v_F$ are involved in the
interband ($\eta_2\neq\eta_3$) and intraband ($\eta_2=\eta_3$) electron-phonon
scattering, separately. Therefore, the negligible thermalization of the
  phonons with $q<\omega_{q\lambda}/v_F$ means that before the establishment of the
hot-electron Fermi distribution ($t\leq131$~fs), the recombination of electrons and
holes due to the interband electron-phonon scattering is marginal. In order to
show this clearer, we plot the $q$ dependence of the angle averaged phonon
temperature ${T}^{\rm ph}_q=\int^{2\pi}_0d\theta_{\bf q}T_{\rm ph}({\bf
  q})/(2\pi)$ at different times in Figs.~\ref{figsw4}(c)-(f), 
as done by Knorr {\it et al.}.\cite{Malic,ButscherAPL91} The locations
  $v_Fq=\omega_{q\lambda}$ for each branch of phonons are indicated by arrows in
  Figs.~\ref{figsw4}(c)-(f). It is found that in the first several hundred femtoseconds,
  only the thermalization of the phonons with $v_Fq>\omega_{q\lambda}$ is
  efficient. After that, the temperatures
    of the phonons with $v_Fq<\omega_{q\lambda}$ increase markedly. Especially,
    for the KO and RI$_2$ branches, the hottest phonons appear among the phonons
    with $v_Fq<\omega_{q\lambda}$ as shown in the inset of Fig.~\ref{figsw4}(f). This is 
consistent with the results by Knorr {\it et al.}\cite{Malic,ButscherAPL91} and can be 
understood as the interband electron-phonon scattering happens only among the
low energy electrons with its energy $|\varepsilon_{\eta{\bf
    k}}|<\omega_{q\lambda}$. However, the energy of the photoexcited electrons
are very high ($\omega_L/2>3.9\omega_{q\lambda}$) and before the low-energy
electrons are heated under the scattering, the energy gained by the electrons
first relaxes through the intraband electron-phonon scattering
process. Then, after the thermalization of the low energy electrons, the
  heating of the phonons with  $v_Fq<\omega_{q\lambda}$ starts to be important.

One also observes from Figs.~\ref{figsw4}(a) and (b) that the hottest
phonons are along the ${\bf q}_y$ direction. This comes from the angular
dependence of the electron-phonon scattering matrices. Since the
hottest phonon appears with $q>\omega_{q\lambda}/v_F$, we only consider the intraband scattering
process. Then, with only the diagonal terms of the density matrices taken into
consideration, the scattering matrices in Eq.~(\ref{epscattering})
$|M^{\mu\mu^\prime\lambda}_{{\bf k},{\bf k-q},\eta\eta}|^2$ are proportional to
$1-\cos(\theta_{\bf k}-\theta_{{\bf k}-{\bf q}})$ for the KO phonons 
and $1+\cos(\theta_{\bf k}-\theta_{{\bf k}-{\bf q}})$ for the RI$_1$
phonons.\cite{Piscanec,Lazzeri,Fratini} One immediately understands that the parallel or 
antiparallel scattering are the strongest. Then, since the photoexcited electrons
are mainly along the ${\bf k}_y$ direction, the hottest phonons, which come from the
electron-phonon scattering, also have to be along the ${\bf y}$ direction as shown
in the figures.  In addition, Figs.~\ref{figsw4}(c)-(f) show that the thermalized
phonons in KO and LO branches occupy  larger momentum range than the other
phonons. This also comes from their different angular
dependences of the scattering 
matrices as revealed by Malic {\it et al.}.\cite{Malic}

\subsection{Low pump-photon energy}
We then investigate the dynamics of electrons and phonons with 
the pump-photon energy $\omega_L$ down to the
THz regime. The maximal pump field in Eq.~(\ref{ElecField})
is taken as $E_0=5$~kV/cm. The substrate is again chosen to be
SiO$_2$.

\begin{figure}[htb]
  \includegraphics[width=8.6cm]{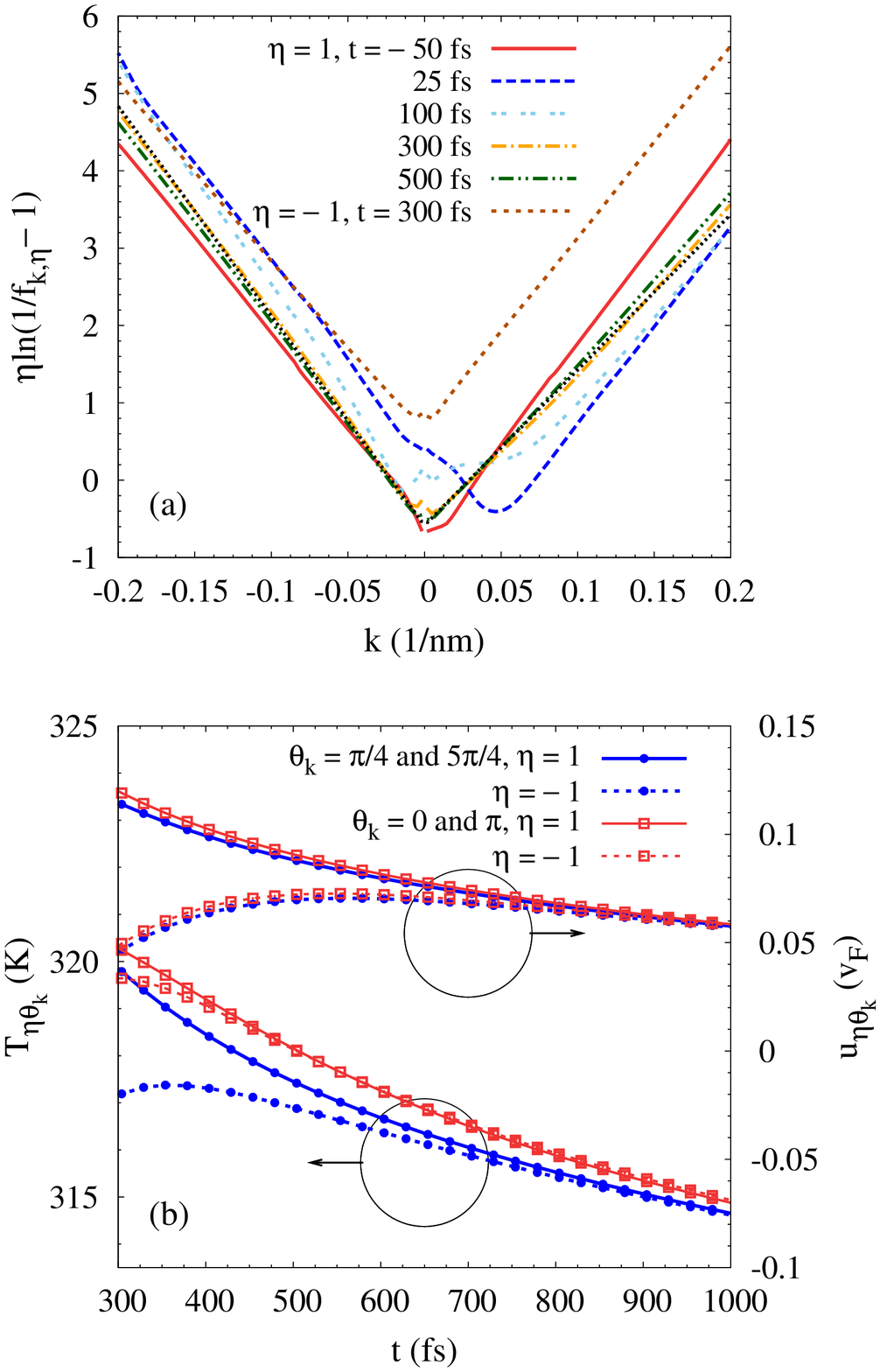}
  \caption{(Color online) (a) $\eta\ln(f^{-1}_{{\bf k},\eta}-1)$
    along $\theta_{\bf k}=0$ and $\pi$ directions for conduction band electrons
    ($\eta=1$) at $t=-50$, 25, 100, 300 and 500~fs, 
    as well as that for valence band electrons ($\eta=-1$) at
    $t=300$~fs. Here, $\omega_L=15$~meV and the impurity density $N_i=0$. The
    black dotted curve is calculated from Eq.~(\ref{DriftFermiDis1}) with $T_1=323.6$~K,
    ${u}_1=0.12v_F$ and $\mu_1=16.3$~meV. 
    (b) Temporal evolution of $T_{\eta\theta_{\bf k}}$ and ${u}_{\eta\theta_{\bf
        k}}$ (with the scale on the right-hand side of the frame) for 
    electrons in conduction (solid curves) and valence (dashed curves) bands obtained by fitting
    Eq.~(\ref{DriftFermiDis1}) along directions $\theta_{\bf k}=0$ and $\pi$ (red
    squares), as well as $\theta_{\bf k}=\pi/4$ and $5\pi/4$ (blue dots).} 
\label{figsw5}
\end{figure}

\subsubsection{Dynamics of electrons}

In this subsection, we investigate the dynamics of electrons with
$\omega_L=15$~meV (3.6~THz). With such low
pump-photon energy, the corresponding oscillation period of the electric field
is 276~fs, which is even larger than twice of the FWHM. Therefore, during
the pump process, the laser field ${\bf E}$ almost keeps  along one direction and
a large net momentum is transferred to electrons. In this case, the drift term
is very important since it describes the momentum transfer. Furthermore, due to
the transferred net momentum, a drifted Fermi distribution after the pulse is
expected. In previous works with static electric 
field,\cite{yzhou,Lifshitz} it has been shown that in the case with strong
Coulomb scattering, the drifted Fermi distribution is expressed as  
\begin{equation}
  f_{{\bf k}\eta}=\{\exp[(\eta v_{\rm F}k-{\bf u}_\eta\cdot{\bf
    k}-\mu_\eta)/(k_BT_\eta)]+1\}^{-1},
  \label{DriftFermiDis1}
\end{equation}
with ${\bf u}_\eta$ being the drift velocity. However, in graphene, it has been
shown by Zhou and Wu that the Coulomb scattering is not strong enough
compared with the drift effect of the static electric field and the drifted
Fermi distribution is instead described by\cite{yzhou} 
\begin{equation}
  f_{{\bf k}\eta}=\{\exp[(\eta v_{\rm F}|{\bf k}-{\bf u}_\eta|-\mu_\eta)/(k_BT_\eta)]+1\}^{-1}. 
  \label{DriftFermiDis2}
\end{equation}
Although both distributions are the same in semiconductors with parabolic
spectrum, they have different properties in graphene due to the linear
dispersion. For Eq.~(\ref{DriftFermiDis2}), one finds that ${\bf
    u}_\eta=\sum_{\bf k}{\bf k}f_{{\bf k},\eta}(t)/\sum_{\bf k}f_{{\bf 
      k},\eta}(t)$ is nothing but the center-of-mass drift velocity. However, in
  Eq.~(\ref{DriftFermiDis1}), ${\bf u}_\eta$ is only an effective parameter. Moreover,
in the $k$ dependence of $\ln(f^{-1}_{{\bf k},1}-1)$ along the direction of ${\bf u}_\eta$, 
Eq.~(\ref{DriftFermiDis1}) indicates a minimum at $k=0$ and the magnitudes of the slopes
for $\theta_{\bf k}$ and $\theta_{\bf k}+\pi$ directions are different. However,
Eq.~(\ref{DriftFermiDis2}) shows that the minimum deviates from $k=0$ and
the magnitudes of the slopes are identical. To reveal the distribution established
here and the dynamics of electrons, we plot $\ln(f^{-1}_{{\bf k},1}-1)$ as functions of $k$ 
along the direction of the electric field ${\bf E}$ (i.e., $\theta_{\bf k}=0$
and $\pi$) at different times in Fig.~\ref{figsw5}(a). One finds that during the
pump process, the minimum of the curve is first drifted away from the Dirac
point as shown by the curves at $t=-50$ and 25~fs. However, before the pulse
ends ($t>100$~fs, when $|E|<0.1|E_0|$), the minimum has already started to tend to
 the Dirac point and the
magnitudes of the slopes for the curves away from Dirac point become
different. Then, at $t=300$~fs, the minimum is around the Dirac point and the
curves along $\theta_{\bf k}=0$ and $\pi$ 
directions are almost linear with $k$, but with different magnitudes of slopes.
With such different magnitudes of slopes, the corresponding drift of the
  center-of-mass $k_{\rm net}$ is 0.02~nm$^{-1}$. This
indicates the establishment of the distribution Eq.~(\ref{DriftFermiDis1}), as
shown by the fitting plotted as the black dotted lines in the figure. This
is understood that the drifted Fermi distribution is established after the pulse in
our present investigation. Then, with the Coulomb scattering 
being the dominant scattering, the distribution as  Eq.~(\ref{DriftFermiDis1})
is established.

It is also shown in Fig.~\ref{figsw5}(a) that at $t=300$~fs, the similar drifted
hot-electron Fermi distribution in the valence band ($\eta=-1$) has also been
established since the minimum of the corresponding curve is around $k=0$ and the curves
along $\theta_{\bf k}=0$ and $\pi$ directions are almost linear with $k$
with different magnitudes of slopes. Moreover, the distributions along the other
directions $\theta_{\bf k}$ are also similar. To further investigate the buildup
of the unitary drifted Fermi distribution along all directions, we fit the
distributions of electrons along different directions in both the conduction and
the valence bands with Eq.~(\ref{DriftFermiDis1}). The obtained 
$T_{\eta\theta_{\bf k}}$ and $u_{\eta\theta_{\bf k}}$ are plotted in
Fig.~\ref{figsw5}(b). One finds that after $t=300$ (380)~fs, the relative
difference between the drift velocities along different directions for electrons
in conduction (valence) band is less than 5~\% and the relative difference of the
temperature increases $T_{1\theta_{\bf k}}-T_0$ ($T_{-1\theta_{\bf k}}-T_0$) is
less than 10~\%. This indicates the buildup of the unitary drifted Fermi
distribution as Eq.~(\ref{DriftFermiDis1}) for all directions in conduction (valence) 
band. It is noted that the case studied here is $n$-doped, which leads to
different buildup times, hot-electron temperatures and drift velocities  
of the unitary drifted Fermi distributions for electrons in conduction and
valence bands. However, under the interband Coulomb scattering, the
temperatures and drift velocities in conduction and valence bands tend to be
identical. From our calculation, the relative difference of the drift velocities 
is less than $10$~\% after $t=650$~fs and that of the temperature increases is
less than $10$~\% after $t=350$~fs, as shown in the figure.

\begin{figure}[htb]
  \includegraphics[width=8cm]{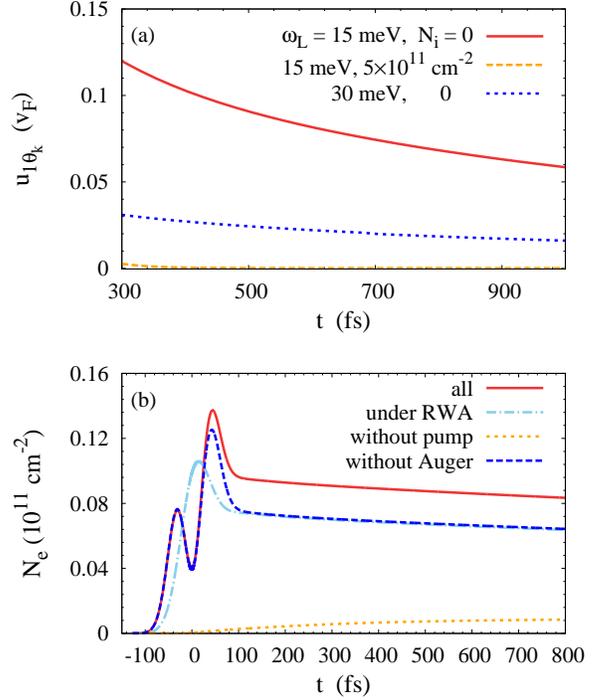}
  \caption{(Color online) (a) Temporal evolution of ${u}_{1\theta_{\bf k}}$
    for the cases with $\omega_L=15$ and 30~meV without impurity, as well as
    $\omega_L=15$~meV with impurity density
    $N_i=5\times10^{11}$~cm$^{-2}$. (b) Temporal evolution of the
photoexcited electron densities for the cases with all the 
processes included, without Auger process,
without pump process and under the rotation-wave approximation (RWA),
separately.
 Here $\omega_L=15$~meV and $N_i=0$.}   
\label{figsw6}
\end{figure}

We also investigate the influence of impurities and pump-photon energy on the
drift velocity of electrons. ${u}_{1\theta_{\bf k}}$ obtained from the
distributions along ${\theta_{\bf k}}=0$ and $\pi$ directions under
different pump-photon energies and impurity densities are 
plotted in Fig.~\ref{figsw6}(a). It is noted that after $t=300$~fs, the relative
differences of the drift velocities along different directions are less than 5~\%
for all cases investigated here. One finds that
in the case without impurity, the drift velocity ${\bf u}_{1\theta_{\bf k}}$
and hence the total momentum relax slowly (with the relaxation time being
about 1.8~ps). However, when the impurities with $N_i=5\times10^{11}$~cm$^{-2}$
(the corresponding mobility is about 5000~cm$^2$/Vs) are introduced into the 
system, the relaxation of the total momentum becomes very fast (with 
the relaxation time being 46~fs) and at $t=300$~fs, the corresponding
${u}_{1\theta_{\bf k}}=0.003v_F$ is negligible. This is because that the
relaxation of the drift velocity comes from the electron-impurity and
electron-phonon scatterings. However, the electron-phonon scattering strength is
weak and only when the impurity density is large (e.g., 
$5\times10^{11}$~cm$^{-2}$), can the relaxation of the net momentum become
fast. Besides, it is also shown in the figure that 
when $\omega_L$ increases from 15 to 30~meV, the value of ${\bf
  u}_{1\theta_{\bf k}}$ at  $t=300$~fs decreases markedly. This is because that
with $\omega_L=30$~meV, the laser field ${\bf E}$ changes direction during the
pulse and the net momentum gained by electrons is partially cancelled out.

We further investigate the influence of the pump and Auger processes on the
evolution of the photoexcited electron density by switching them off
separately. The results are plotted in Fig.~\ref{figsw6}(b).
In the calculation, $\omega_L=15$~meV and $N_i$=0. It is
noted that without the pump process, electrons can 
still be excited from the valence band to conduction band. This is because that
the electrons are heated by the drift effect of the laser field, which leads to the
interband electron-phonon scattering. However, one finds from the figure that the dominant
mechanism for the excitation of the electrons is the pump process in the case
investigated here.  Moreover, it is also found that the Auger process indeed
affects the dynamics of electrons and it leads to the change of the excited
electron density at the end of the pulse ($t=100$~fs)
 as large as 22~\%, which is
consistent with our analysis in Sec.~{\Rmnum 2}. Furthermore, with such low
  pump-photon energy, the widely accepted rotation-wave approximation in
semiconductors optics is no longer valid. The temporal 
evolution of the photoexcited electron density under the 
rotation wave approximation is also plotted in
Fig.~\ref{figsw6}(b). One finds that the fast oscillation of the electron
 density during the pump process disappears and the excited electron density at
  the end of the pulse changes about 17~\%.

\begin{figure}[htp]
  \includegraphics[width=7.7cm]{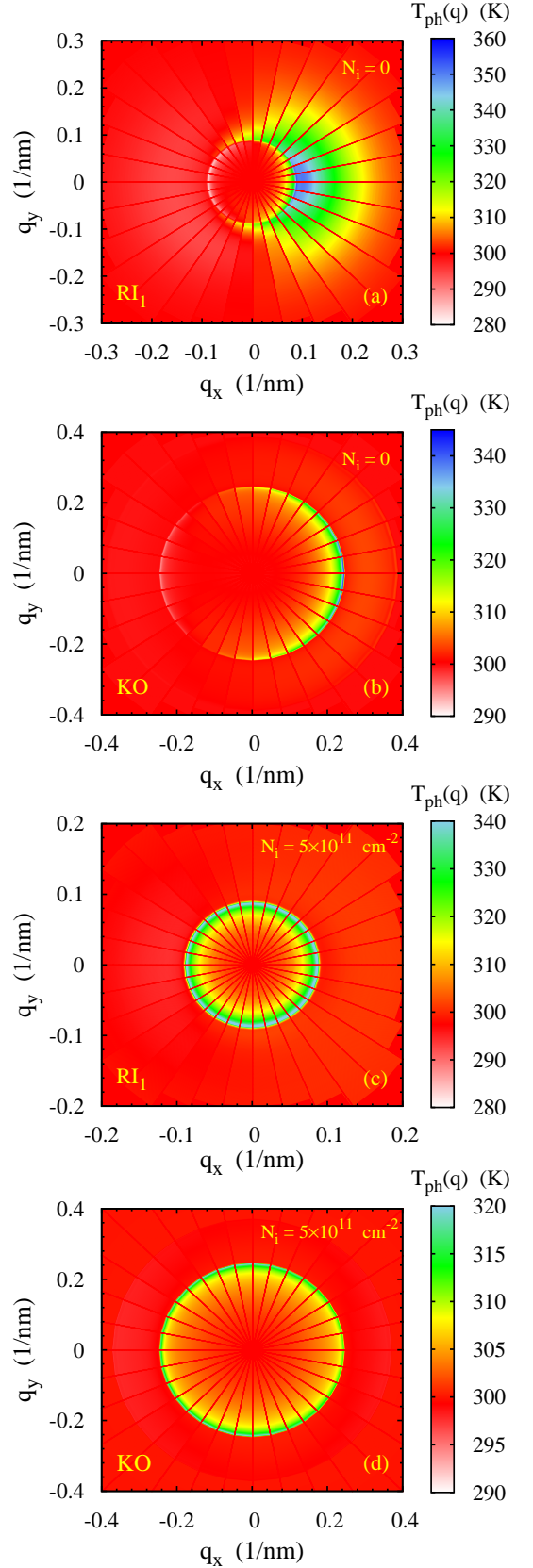}
  \caption{(Color online) (a) and (b) [(c) and (d)] ${\bf q}$-resolved
    hot-phonon temperatures with $N_i=0$ ($5\times10^{11}$)~cm$^{-2}$ at
    $t=300$~fs for RI$_1$ and KO phonons, respectively.} 
\label{figsw7}
\end{figure}

\subsubsection{Dynamics of phonons}
We further investigate the dynamics of phonons. We plot the ${\bf
  q}$-resolved temperatures of the hot RI$_1$ and KO phonons with
$\omega_L=15$~meV under the impurity densities $N_i=0$ and $5\times10^{11}$~cm$^{-2}$ in
Fig.~\ref{figsw7}.
The other phonons are similar and not shown. The temperatures of these two
branches of phonons are obtained at $t=300$~fs when the unitary distribution of
 electrons for all directions has been established. From the figures, one
finds that in the case without impurity, the ${\bf q}$-resolved temperatures of
the hot phonons are highly anisotropic. Moreover, by considering that ${\bf
  u}_\eta$ in both conduction and valence bands are along the $\theta_{\bf k}=0$
direction, it is shown that the hottest phonons appear in the direction parallel 
to ${\bf u}_\eta$ and the coldest phonons antiparallel to ${\bf u}_\eta$. This
phenomenon is also shown in semiconductors under static electric
field.\cite{Zhang2010} It can be understood from the generation rate of phonons due to
the electron-phonon scattering [Eq.~(\ref{nph})] after the
establishment of the drifted Fermi distribution of electrons. With only the
diagonal terms of the density matrices taken into consideration, Eq.~(\ref{nph})
is given by 
\begin{eqnarray}
&&\hspace{-0.6 cm}\left.\partial_t n^\lambda_{\bf q}\right|_{\rm ep}=2\pi\sum_{{\bf
    k}\eta_1\eta_2\atop\mu\mu^\prime}|M^{\mu\mu^\prime}_{{\bf k+q},{\bf
    k},\eta_2\eta_1}|^2\delta(\varepsilon_{\eta_2{\bf k+q}}-\varepsilon_{\eta_1{\bf
    k}}-\omega_{q\lambda})\nonumber \\
&&\hspace{0. cm}\mbox{}\times (f_{\mu^\prime{\bf k},\eta_1}^>{f}_{\mu{\bf
    k+q},\eta_2}^<N^{+\lambda}_{\bf q}-{f}_{\mu^\prime{\bf k},\eta_1}^<{f}_{\mu{\bf
    k+q},\eta_2}^>N^{-\lambda}_{-{\bf q}}).
\end{eqnarray}
Substituting the drifted Fermi distribution Eq.~(\ref{DriftFermiDis1}) and the
Bose distribution of phonons $n^{\lambda}_{\bf
  q}=1/\{\exp[\omega_{q\lambda}/(k_BT_{\rm ph})]-1\}$ into it, one finds that whether
the phonons are emitted or absorbed is determined by the sign of the term 
\begin{equation}
e^{(\eta_1 v_{\rm
  F}k-{\bf u}\cdot{\bf k}-\mu+\omega_{q\lambda})/(k_BT)}-e^{[\eta_2 v_{\rm
  F}|{\bf k}+{\bf q}|-{\bf u}\cdot({\bf k}+{\bf q})-\mu]/(k_BT)}. 
\label{phononunderdrift}
\end{equation}
Here, to simplify the analysis, the temperatures of phonons and electrons are taken to be
the same $T$ and the chemical potentials and drift velocities for electrons in
different bands are also set to be unified ${\mu}$ and ${\bf u}$,
respectively. Then, due to the energy conservation
$\delta(\varepsilon_{\eta_2{\bf k+q}}-\varepsilon_{\eta_1{\bf 
    k}}-\omega_{q\lambda})$, Eq.~(\ref{phononunderdrift}) is simplified to $e^{(\eta_1 v_{\rm
  F}k-{\bf u}\cdot{\bf k}-\mu+\omega_{q\lambda})/(k_BT)}[1-e^{-{\bf u}\cdot{\bf
  q}/(k_BT)}]$. One immediately understands that the hottest phonons appear
among phonons with ${\bf u}$ parallel to ${\bf q}$ and the coldest phonons appear
among the phonons with ${\bf u}$ antiparallel to ${\bf q}$ as shown in Figs.~\ref{figsw7}(a)
and (b). This phenomenon can also be understood physically from the
relaxation of the drift velocity of electrons.  During the relaxation, the net
momentum gained from the laser field is transferred from the electrons to the
phonons, which implies the tendency of the absorption of the phonons with ${\bf q}\cdot{\bf
  u}_\eta<0$ and the emission of the phonons with ${\bf q}\cdot{\bf u}_\eta>0$. On
the other hand, in the system where the net momentum relaxes very fast through the
electron-impurity scattering, the distributions of phonons become more isotropic as shown
in Figs.~\ref{figsw7}(c) and (d).

\subsection{Negative DT in system with high Fermi energy}
In this subsection, we further investigate the negative DT due to the weakening of the
Pauli blocking. Sun {\it et al.} measured the DT with high
probe-photon energy ($>500$~meV) and showed that a peak exists in the temporal
evolution of DT.\cite{DSunPRL} Moreover, it is also found that there is a range of the 
probe-photon energy where the negative DT appears during the evolution. To
understand this, they suggested that their epitaxial sample consists of a stack of
graphene layers\cite{Dawlaty,Latil,Varchon} with some layers heavily doped and some layers undoped. Moreover, the Fermi
energy of the heavily doped layers is assumed to be around half of the upper boundary of the
probe energy range for the appearance of the negative DT. Then, with the energies
of the resonant absorption states of the probe pulse smaller than the Fermi
energy in the heavily doped layers, the corresponding distribution difference
$f_{{\bf k_\omega},1}-f_{{\bf k_\omega},-1}$ in Eq.~(\ref{conductivity})
increases in the undoped layers but decreases in the
heavily doped layers after the pulse. This leads to the decrease of the
conductivity in undoped layers but increase in heavily doped
ones. These two opposite tendencies compete in the evolution and are
responsible for the peak and the negative DT. Moreover, with the
decrease of the probe-photon energy, the variation of the conductivity from the
undoped layers is strengthened whereas that from the heavily doped layers
is weakened. Furthermore, if the probe-photon energy is higher than twice of the Fermi
energy in the heavily doped layers, the conductivities in both the heavily and
undoped layers decrease. These properties lead to the fact that the negative DT
appears only in a probe-energy range smaller than twice of the Fermi energy.

\begin{figure}[htb]
  \includegraphics[width=7.5cm]{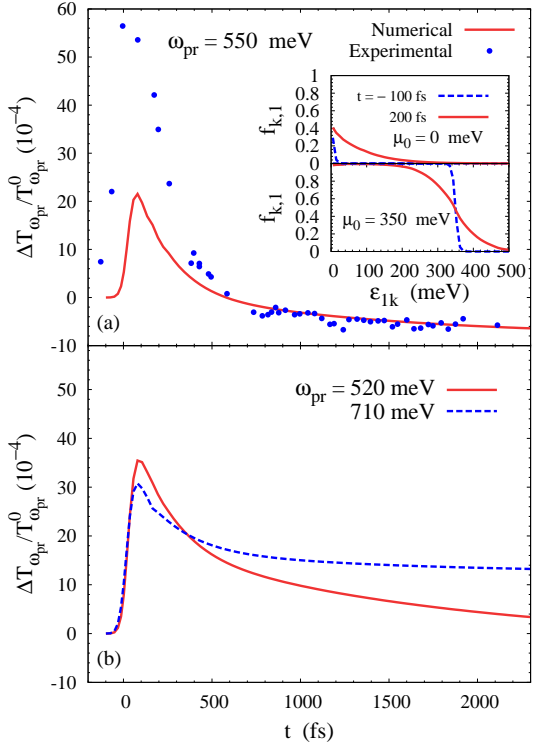}
  \caption{(Color online) (a) DT from the numerical results compared with the
    experimental data extracted from Fig. 3(a) in Ref.~\onlinecite{DSunPRL}. The inset shows the electron distributions before ($t=-100$~fs) and
    after ($t=200$~fs) the pulse in the heavily doped and undoped layers.
    (b) Temporal evolution of DT under different probe-photon energies.} 
\label{figsw8}
\end{figure}

To check above conjecture, we fit their experimental data with our microscopic approach. The
substrate is chosen to be SiC, the environmental temperature is 50~K, the FWHM
of the pulse is 100~fs and the pump-photon energy $\omega_L=1550$~meV as
indicated by Sun {\it et al.}.\cite{DSunPRL} The maximal electric field in 
Eq.~(\ref{ElecField}) is taken to be $E_0=175$~kV/cm. With such pump intensity,
the photoexcited electron density is about $4\times10^{11}$~cm$^{-2}$ per
layer. The probe-photon energy $\omega_{\rm pr}=550$~meV is the same as that in
the experiment. It is noted that with such high probe energy, the intraband  absorption
is suppressed.\cite{Scharf} To model the multilayer structure in their sample, 
$N_{\rm lay}\sigma_{\omega_{\rm pr}}(t)$ in Eq.~(\ref{Trans}) is replaced by
$N_{h}\sigma_{h}(t)+N_{u}\sigma_{u}(t)$ with $\sigma_{h}(t)$ and $\sigma_{u}(t)$
being the conductivity in heavily and undoped graphenes, respectively and
$N_h$ and $N_u$ being the corresponding layer numbers. 
In our computation, $\sigma_{h}(t)$ and $\sigma_{u}(t)$ are calculated
separately with the corresponding equilibrium Fermi energy being $\mu_0=350$ and
$0$~meV, respectively, as estimated in the experiment. $N_h=2$ and $N_u=20$ are
taken as fitting parameters which are very close to those estimated in the
experiment. The results are plotted in 
Fig.~\ref{figsw8}(a). One finds that our numerical results repeat the peak and
the negative DT shown by the experimental data. Especially, when $t>500$~fs, the
fitting is quite well. The discrepancy between our numerical results with the
experimental data for the magnitude of the peak is again from the
approximation mentioned in Sec.~IIIA2 as well as the simplification of the
complex layer configuration to two kinds of doped layers. The electron
distributions before and after the pulse are plotted in the inset of
Fig.~\ref{figsw8}(a), which is consistent with the arguments by Sun {\it et
  al.}.\cite{DSunPRL} Moreover, when the probe-photon energy is small enough or larger than
twice of the Fermi energy, the DT keeps positive
as shown by the curves with $\omega_{\rm pr}=520$ and 710~meV in
Fig.~\ref{figsw8}(b).  Based on these results, our microscopic investigation
supports their argument that the negative DT comes from the weakening of the
Pauli blocking in the heavily doped layers. 

\section{SUMMARY}
In summary, we have investigated the nonequilibrium dynamics of carriers and
phonons in graphene under the linearly
  polarized near-infrared and THz laser pulses via the microscopic 
kinetic Bloch equation approach.\cite{wuReview,Haug} The carrier-impurity,
carrier-phonon and carrier-carrier Coulomb scatterings are explicitly included
and the dynamic screening for the Coulomb scattering is utilized. Moreover,
based on the vector potential gauge and the gauge invariant approach,\cite{Haug}
both the drift and pump terms are naturally included in our approach.

We first investigate the thermalization of electrons due to the drift term. It
is shown that the thermalization is weakened with the increase of
the pump-photon energy and it is even negligible when the pump-photon energy is
higher than 500~meV. Moreover, a peak appears in the impurity density dependence of the
thermalization. Then, the dynamics of carriers and phonons 
with the pump-photon energy up to the near-infrared regime is studied. We show that the
temporal evolution of the DT reported by Hale {\it et al.}\cite{Hale} can be well 
fitted. Moreover, under the linearly polarized laser pulse, the electrons
  are photoexcited anisotropically. However, their distribution tends to be isotropic
 under the scattering. As a result, 
the isotropic hot-electron Fermi distribution is found to be established within
  $t=131$~fs, which is before the end of the pulse 
(the FWHM of the pulse is 180~fs). This is consistent with the previous
works.\cite{Sun,Malic,BreusingPRB83} In contrast to the high pump-photon energy
case, in the case with  the pump-photon energy down to the terahertz regime, we show that a
large net momentum is transferred to electrons through 
the drift term and a drifted Fermi distribution different from the one under
static field is established in several hundred femtoseconds.\cite{yzhou} We further
show that the Auger process investigated in the previous
works\cite{Winzer,Kim,WinzerPRB85} which only involves the diagonal terms of
density matrices is forbidden by the dynamic screening. However, the Auger
process involving the interband coherence contributes to the 
dynamics of carriers. Nevertheless, its influence is important only when the pump-photon
energy is low. In contrast, for the rotation-wave approximation which
is widely accepted in semiconductor optics, we show that it fails when
the pump energy is low. We also apply our microscopic theory to investigate the negative
DT in the case where the equilibrium Fermi energy is high and the probe-photon
energy is around twice of the Fermi energy. We show that the negative DT appears
due to the weakening of the Pauli blocking under the pump pulse, which supports
the suggestion by Sun {\it et al.} in their experimental work.\cite{DSunPRL}

In addition, the momentum-resolved hot-phonon temperatures are also studied. When the
pump-photon energy is high, the hot phonon temperatures are anisotropic due to
the linearly polarized laser field and the hottest phonons appear with their momentum in the direction  
perpendicular to the laser field ${\bf E}$. Moreover, we also show that the
electron-hole recombination due to the interband  
electron-phonon scattering is marginal in the first several hundred
femtoseconds. On the other hand, when the 
pump-photon energy is low and the relaxation of the net momentum is slow, the
hot-phonon temperatures are also highly anisotropic with the hottest and the
coldest phonons appearing in the directions parallel and antiparallel to the
drift velocity, respectively, due to the drifted Fermi
distribution of electrons. However, if the relaxation of the net momentum is
fast as in the case with high impurity density, the hot-phonon temperatures
become more isotropic.

\begin{acknowledgments}
  The authors would like to thank M. Q. Weng for valuable discussions on the numerical scheme
  of the drift term. This work was supported by the National Basic Research 
Program of China under Grant No. 2012CB922002 and the Strategic Priority 
Research Program of the Chinese Academy of Sciences under Grant
No. XDB01000000. 
\end{acknowledgments}

\appendix
\section{Gauge Invariant Kinetic Bloch Equations}
We use the nonequilibrium Green function approach to derive the gauge invariant
kinetic Bloch equations under the optical field.\cite{wuReview,Haug} 
The effective Hamiltonian without external field in the real space is given by\cite{PeresPRB}
\begin{equation}
H^{\mu,{\rm AB}}_{0,{\rm r}}(-i\nabla)=v_F(-i\mu\tau_x\nabla_x-i\tau_y\nabla_y),
\end{equation}
and the electron--optical-field interaction Hamiltonian is described by\cite{PeresPRB}
\begin{equation}
H^{\mu,{\rm AB}}_{\rm photon,r}(t)=|e|v_F[\mu\tau_xA_x(t)+\tau_yA_y(t)]/c.
\end{equation}
Then, in the base set of the eigenstates of $H^\mu_{0,{\rm r}}(-i\nabla)$, the above
Hamiltonians read
\begin{eqnarray}
&&\hspace{-1.1cm}H^\mu_{0,{\rm r}}({\bf p})=v_Fp\sigma_z,\\
&&\hspace{-1.1cm}H^\mu_{\rm photon,r}({\bf p},t)={|e|v_F}{\bf A}(t)\cdot[{\bf p}\sigma_z+\mu \hat{\bf z}\times{\bf p} \sigma_y]/(cp),
\end{eqnarray}
with $p_x=-i\partial_x$, $p_y=-i\partial_y$ and $p=\sqrt{p^2_x+p^2_y}$. It is
noted that this electron--optical-field interaction Hamiltonian is consistent
with the previous works based on the tight-binding approach.\cite{Stroucken2011,Malic2006}

The contour-ordered Green function $G^\mu(1,2)$ are matrices with the
matrix elements defined as $G^\mu_{\eta\eta^\prime}(1,2)=-i\langle
T_c[\psi^{\mu}_{\eta}(1)\psi^{\mu\dag}_{\eta^\prime}(2)]\rangle$. Here $\psi^{\mu}_{\eta}$
and $\psi^{\mu\dag}_{\eta}$ are the Fermion field operators in the Heisenberg picture
for electrons in band $\eta$ and valley $\mu$; $(1)=({\bf r_1},t_1)$ and $(2)=({\bf r}_2,t_2)$
stand for the space-time points on the contour and $G^\mu_0$
represent the free-particle Green function. By multiplying
$i\overrightarrow{\partial}_{t_1}-H^\mu_{0,{\rm r}}(\overrightarrow{\bf
p}_1)$ and $-i\overleftarrow{\partial}_{t_2}-H^\mu_{0,{\rm r}}(-\overleftarrow{\bf p}_2)$
to the Dyson equation of $G^\mu(1,2)$ and
subtracting the obtained two equations from each other, one obtains the equation
with ${\bf r_1}$, $t_1$, ${\bf r_2}$ and $t_2$. Then, transforming the equation
to the center-of-mass and relative variables ${\bf R}=({\bf r_1}+{\bf r_2})/2$,
$T=({t_1+t_2})/{2}$, ${\bf r}={\bf r_1}-{\bf r_2}$ and $\tau=t_1-t_2$, and
omitting ${\bf R}$ due to the uniformity of the system studied here, one
obtains\cite{Haug} 
\begin{eqnarray}
&&i{\partial}_TG^\mu(\tau,T,{\bf r})=[H^\mu_{0,{\rm r}}({\bf p}_r)+H^\mu_{\rm photon,r}({\bf p}_r,T+\tau/2)]\nonumber \\
&&\hspace{0.75cm}\mbox{}\times G^\mu(\tau,T,{\bf r})-G^\mu(\tau,T,{\bf
  r})[H_{0,{\rm r}}({\overleftarrow{\bf p}_r})\nonumber \\
&&\hspace{0.75cm}\mbox{}+H^\mu_{{\rm photon,r}}(\overleftarrow{\bf p}_r,T-\tau/2)].
\label{EqCenterRelative}
\end{eqnarray}
Here ${\bf p}_r$ represents the momentum operator with
respect to the relative variable. The gauge
invariant functions are constructed as\cite{Haug}
\begin{equation}
G^\mu({\bf k},\tau,T)=\int d^2r\exp({-i{\bf \bar{k}\cdot r}})G^\mu(\tau,T,{\bf
  r}),
\end{equation}
with ${\bf \bar{k}}={\bf k}-\int^{1/2}_{-1/2}d\lambda {|e|}{\bf
  A}(T+\lambda \tau)/c$. Applying this transformation to
Eq.~(\ref{EqCenterRelative}) and setting $\tau=0^-$, one has
\begin{eqnarray}
&\hspace{-2cm}{\partial}_T{\rho}_{\mu{\bf k}}(T)+{|e|}{\partial_T {\bf A}(T)}\cdot\nabla_{{\bf k}}{\rho}_{\mu{\bf k}}(T)/c\nonumber\\
&\mbox{}+i[H^\mu_{0,{\rm r}}(\bar{\bf k})+H^\mu_{\rm photon,r}(\bar{\bf k},T),{\rho}_{\mu{\bf k}}(T)]=0,
\label{KEqusiSpin}
\end{eqnarray}
with ${\rho}_{\mu{\bf k}}(T)=-iG^\mu({\bf k},0^-,T)$ being the local density
matrix of electrons with momentum ${\bf k}$ and $H^\mu_{\rm photon,r}(\bar{\bf
  k},T)$ coming from the electron--optical-field interaction given by
Eq.~(\ref{PhotonElec_H}). ${|e|}{\partial_T {\bf
    A}(T)}\cdot\nabla_{{\bf k}}{\rho}_{\mu{\bf k}}(T)/c$ comes from the
transformation of ${\partial}_TG^\mu(\tau,T,{\bf r})$ and is just the drift
term since ${\bf E}(T)=-{\partial_T {\bf A}(T)/c}$ under the vector potential
gauge. Moreover, the derivation here is based on the assumption that the field
${\bf A}$ is small and can be treated perturbatively. Then, expanding $|\bar{\bf
  k}|$ around $|{\bf k}|$ and keeping only the first order term of ${\bf A}(t)$,
one  obtains Eq.~(\ref{KEE}) except for the scattering term and Hartree-Fock
term.

\section{Expression of the scattering terms}
In this section, we give the scattering matrices used in
Eqs.~(\ref{eiscattering}) and (\ref{epscattering}). $U(\varepsilon_{\eta_2{\bf
    k}}-\varepsilon_{\eta_2{\bf k-q}})=\sqrt{N_i}Z_iV^r(\varepsilon_{\eta_2{\bf
    k}}-\varepsilon_{\eta_2{\bf k-q}})\exp({-qd})$ with $Z_i = 1$ being the charge
number of the impurity, $N_i$ standing for the impurity density and $d$
representing the effective distance of the 
impurity layer to the graphene sheet. $M^{\mu^\prime\mu\lambda}_{{\bf
    k^\prime,k},\eta\eta^\prime}$ are scattering matrices for electron-phonon scattering with
phonons in branch $\lambda$. For the AC phonon, $\omega_{{\bf 
    q}{\rm AC}}=v_{\rm ph}q$ with $v_{\rm ph}$ being the sound velocity and
the corresponding scattering matrices are
\begin{equation}
M^{{\rm AC}\mu^\prime\mu}_{\bf k^\prime\eta^\prime,k\eta}={D\sqrt{q}}S_{{\bf
    k^\prime}\eta^\prime,{\bf k\eta}}\delta_{\mu^\prime\mu}/{\sqrt{2\rho_mv_{\rm
      ph}}},
\end{equation}
in which $D$ is the deformation potential and $\rho_m$ denotes the graphene mass
density.\cite{Chen,Hwang3} For the RI phonons,
\begin{equation}
M^{{\rm RI}\mu^\prime\mu}_{\bf k^\prime\eta^\prime,k\eta}=\sqrt{g}
   {v_F e^{-qd}}{S_{{\bf k^\prime}\eta^\prime,{\bf k\eta}}\delta_{\mu\mu^\prime}}/{\sqrt{a q \epsilon^{\rm RI}_{\bf q}}},\label{MRI} 
\end{equation}
where $g$ represents the dimensionless coupling parameter depending on the
material of the substrate,\cite{Perebeinos,Fratini} $a$ is
the C-C bond distance and $\epsilon^{\rm RI}_{\bf q}$ is the substrate-dependent
dielectric function. For the two branches of phonons RI$_1$ and RI$_2$
in the sample on SiO$_2$, the corresponding dielectric functions are\cite{Fischetti2001}
\begin{eqnarray}
1/{\epsilon^{\rm RI_1}_{\bf q}}=1/[{\epsilon_i+\epsilon({\bf
    q},0)}]-1/[{\epsilon_0+\epsilon({\bf q},0)}],\\
1/{\epsilon^{\rm RI_2}_{\bf q}}=1/[{\epsilon_\infty+\epsilon({\bf
    q},0)}]-1/[{\epsilon_i+\epsilon({\bf q},0)}],
\end{eqnarray} 
with $\epsilon_i=3.36$, $\epsilon_0=3.90$ and
$\epsilon_\infty=2.40$.\cite{Perebeinos} For the only branch of RI phonons in graphene on SiC
substrate, it is\cite{Fischetti2001}
\begin{equation}
1/\epsilon^{\rm RI}_{\bf q}=1/[\epsilon_\infty+\epsilon({\bf
    q},0)]-1/[\epsilon_0+\epsilon({\bf q},0)],
\end{equation}
with $\epsilon_0=9.7$ and $\epsilon_\infty=6.5$.\cite{Perebeinos,Harris1995}

For the optical phonons, the scattering matrices are
obtained according to the work by Ishikawa {\it et al.}\cite{Ishikawa} as
\begin{equation}
M^{{\rm OP}\mu^\prime\mu}_{\bf k^\prime\eta^\prime,k\eta}={A^{\rm
    OP}_{\mu^\prime\mu\eta^\prime\eta{\bf k^\prime k}}}/{\sqrt{\rho_m\omega_{\rm OP}}},\label{MOP}
\end{equation}
with the corresponding parameters of the optical phonons given by 
\begin{eqnarray}
&&\hspace{-0.65cm}A^{\rm  LO}_{\mu^\prime\mu\eta^\prime\eta{\bf k+q},{\bf k}}={i\delta_{\mu\mu^\prime}}\sqrt{\langle
D^2_{\rm \Gamma}\rangle}[\eta e^{i\mu(\theta_{\bf q}-\theta_{\bf k})}\nonumber
\\
&&\mbox{}-\eta^\prime
e^{i\mu(\theta_{\bf k+q}-\theta_{\bf q})}]/2,\label{ALO} \\ 
&&\hspace{-0.65cm}A^{\rm TO}_{\mu^\prime\mu\eta^\prime\eta{\bf k+q},{\bf k}}={-\mu\delta_{\mu\mu^\prime}}\sqrt{\langle
D^2_{\rm \Gamma}\rangle}[\eta e^{i\mu(\theta_{\bf q}-\theta_{\bf k})}\nonumber\\
&&\hspace{-0.cm}\mbox{} +\eta^\prime e^{i\mu(\theta_{\bf k+q}-\theta_{\bf q})}]/2,\label{ATO}
 \\
&&\hspace{-0.65cm}A^{\rm KO}_{\mu^\prime\mu\eta^\prime\eta{\bf k+q},{\bf k}}={\sqrt{\langle
D^2_{\rm K}\rangle}}[\mu\eta e^{-i\mu\theta_{\bf k}}\nonumber \\
&&\mbox{}+\mu^\prime\eta^\prime
e^{i\mu^\prime\theta_{\bf k+q}}]\delta_{\mu,-\mu^\prime}/2.\label{KTO}
\end{eqnarray}

\section{Numerical scheme for drift term}
In this section, we present the numerical scheme for the drift term. The
truncated 2D momentum space is divided into $N\times M$ control
regions in polar coordinate system,\cite{Weng} each with equal energy and angle
intervals. The grid point ${\bf k}_{n,m}$ is localed in the center of the control
region $(n,m)$ with its radial coordinate $k_n=(n+0.5)\Delta k$ and angular
coordinate $\theta_m=m\Delta \theta$. Here, $\Delta k$ is the momentum interval
and $\Delta \theta=2\pi/M$ is the angular interval. 

The drift term is dealt with the discrete conservation
principle:
\begin{eqnarray}
&&\hspace{-0.45cm}\left.|e|{\bf E}(t)\cdot{\nabla}_{{\bf k}}
  \rho_{\mu{\bf k},\eta\eta^\prime}\right|_{{\bf k}={\bf k}_{n,m}} \simeq
{\int_{\Omega_{n,m}} d^2k \;
  \frac{|e| E(t) \partial_{k_x}  \rho_{\mu{\bf k},\eta\eta^\prime}}{ k \Delta k\Delta \theta}} \nonumber \\
&&\hspace{-0.45cm} = \frac{|e| E(t) }{k \Delta k\Delta \theta}
\int_{\Omega_{n,m}} dk d\theta_{\bf k}  \; [\partial_k
(k\rho_{\mu{\bf k},\eta\eta^\prime}\cos\theta_{\bf k}) \nonumber \\ 
&& \hspace{-0.25 cm} \mbox{}-\partial_{\theta_{\bf  k}}(\rho_{\mu{\bf k},\eta\eta^\prime}\sin\theta_{\bf k})]
 \nonumber \\
&&\hspace{-0.45cm} \simeq \frac{|e|E(t)}{ k \Delta k \Delta \theta}
[\Delta\theta (F^{r,n^\prime m^\prime}_{\mu,nm,\eta\eta^\prime}-F^{r,n^\prime m^\prime}_{\mu,n-1,m,\eta\eta^\prime})\nonumber\\
&&\hspace{-0.25 cm} \mbox{}-\Delta k(\sin{\theta_{m+0.5}}F^{\theta,m^\prime}_{\mu,nm,\eta\eta^\prime}-\sin{\theta_{m-0.5}}F^{\theta,m^\prime}_{\mu,n,m-1,\eta\eta^\prime})].\nonumber\\
\end{eqnarray}
Here $\Omega_{n,m}$ and $\partial \Omega_{n,m}$ are the 
control region  which
contains the grid point ${\bf k}_{n,m}$ and the corresponding
boundary, respectively, and the electric field is set to be along the ${\bf x}$
direction. In the last step of the above equation, the integration of the 
boundary is replaced by the summation over the first order quadrature
on the four (or three if the control region is the neighbor of
${\bf k}=0$) sides of the boundary $\partial\Omega_{n,m}$ and the fluxes
   treated up to third order\cite{Carrillo2003} in the polar coordinate are given by 
\begin{eqnarray}
&&\hspace{-1.05cm}F^{r,n^\prime
  m^\prime}_{\mu,nm,\eta\eta^\prime}=-{k_{n^\prime-1}}\cos\theta_{m^\prime}\rho_{\mu{\bf
    k}_{n^\prime-1,m^\prime},\eta\eta^\prime}/6\nonumber \\
&&\hspace{-0.75cm}\mbox{}+\cos\theta_m({5k_{n^\prime}}\rho_{\mu{\bf
    k}_{n^\prime,m},\eta\eta^\prime}+{2k_{n^\prime+1}}\rho_{\mu{\bf
    k}_{{n^\prime+1},m},\eta\eta^\prime})/6,\\
&&\hspace{-1.05cm}F^{\theta,m^\prime}_{\mu,nm,\eta\eta^\prime}=5\rho_{\mu{\bf
    k}_{n,m^\prime},\eta\eta^\prime}/6+\rho_{\mu{\bf
    k}_{n,m^\prime+1},\eta\eta^\prime}/3\nonumber \\
&&\hspace{-0.75cm}\mbox{}-\rho_{\mu{\bf k}_{n,m^\prime-1},\eta\eta^\prime}/6.
\end{eqnarray}
In order to satisfy the requirement of the numerical stability, $m^\prime$ in
$F^{\theta,m^\prime}_{\mu,nm,\eta\eta^\prime}$ is taken to be $m$ if $-|e|{\bf
  E}(t)\cdot\hat{\bf x}\sin\theta_{m+0.5} > 0$ and $m+1$ otherwise. Moreover, $n^\prime$ in
$F^{r,n^\prime m^\prime}_{\mu,nm,\eta\eta^\prime}$ is chosen to be $n$ if $-|e|{\bf
  E}(t)\cdot\hat{\bf x}\cos\theta_m > 0$ and $n+1$ otherwise. Besides, if
$n^\prime-1\geq 0$, $m^\prime$ is taken to be $m$. On the other hand, if $n^\prime-1<0$, the term
$n^\prime-1$ in the equation is replaced by $0$ and $m^\prime=m+M/2$. It is
noted that with this numerical scheme, the temperature variation due to the
noise is within 3~K for the pulse investigated in Fig.~\ref{figsw2}.


\begin{thebibliography}{0}


\bibitem{Avouris} P. Avouris, Z. Chen, and V. Perebeinos, Nat. Nanotech.
  {\bf 2}, 605 (2007).

\bibitem{NetoReview}  A. H. Castro Neto, F. Guinea, N. M. R. Peres, K. S. Novoselov, and
  A. K. Geim, Rev. Mod. Phys. {\bf 81}, 109 (2009).

\bibitem{Peres} N. M. R. Peres, Rev. Mod. Phys. {\bf 82}, 2673 (2010).

\bibitem{ChoiCRSSMS} W. Choi, I. Lahiria, R. Seelaboyinaa, and Y. S. Kangb, Crit. Rev. Solid State Mater. Sci. {\bf 35}, 52 (2010).

\bibitem{Schwierz} F. Schwierz, Nat. Nanotech. {\bf 5}, 487 (2010).

\bibitem{AvourisNanoLett} P. Avouris, Nano Lett. {\bf 10}, 4285 (2010).

\bibitem{AbergelAdvPhy} D. S. L. Abergel, V. Apalkov, J. Berashevich,
  K. Ziegler, and T. Chakraborty, Adv. Phys. {\bf 59}, 261 (2010).

\bibitem{Bonaccorso} F. Bonaccorso, Z. Sun, T. Hasan, and A. C. Ferrari,
  Nat. Photon. {\bf 4}, 611 (2010).

\bibitem{DSarmaReview} S. Das Sarma, S. Adam, E. H. Hwang, and E. Rossi,
  Rev. Mod. Phys. {\bf 83}, 407 (2011).

\bibitem{Young} A. F. Young and P. Kim, Annu. Rev. Condens. Matter Phys. {\bf 2}, 101 (2011).

\bibitem{Castro} A. H. Castro Neto and K. S. Novoselov, Rep. Prog. Phys. {\bf 74}, 082501 (2011).

\bibitem{Kotov} V. N. Kotov, B. Uchoa, V. M. Pereira, F. Guinea, and A. H. Castro Neto, Rev. Mod. Phys. {\bf 84}, 1067 (2012).




\bibitem{Dawlaty} J. M. Dawlaty, S. Shivaraman, M. Chandrashekhar, F. Rana, and
  M. G. Spencer, Appl. Phys. Lett. {\bf 92}, 042116 (2008).

\bibitem{DSunPRL} D. Sun, Z.-K. Wu, C. Divin, X. Li, C. Berger, W. A. de Heer,
P. N. First, and T. B. Norris, Phys. Rev. Lett. {\bf 101}, 157402 (2008).

\bibitem{GeorgeNanoLett} P. A. George, J. Strait, J. Dawlaty, S. Shivaraman,
  M. Chandrashekhar, F. Rana, and M. G. Spencer, Nano Lett. {\bf 8}, 4248 (2008).

\bibitem{ChoiApl94} H. Choi, F. Borondics, D. A. Siegel, S. Y. Zhou, M. C. Martin,
  A. Lanzara, and R. A. Kaindl, Appl. Phys. Lett. {\bf 94}, 172102 (2009).

\bibitem{NewsonOpt} R. W. Newson, J. Dean, B. Schmidt, and H. M. van Driel, Opt. Express {\bf 17}, 2326 (2009).


\bibitem{BreusingPRL102} M. Breusing, C. Ropers, and T. Elsaesser,
  Phys. Rev. Lett. {\bf 102}, 086809 (2009).

\bibitem{Plochocka} P. Plochocka, P. Kossacki, A. Golnik, T. Kazimierczuk, C. Berger,
  W. A. de Heer, and M. Potemski, Phys. Rev. B {\bf 80}, 245415 (2009).

\bibitem{Wang} H. Wang, J. H. Strait, P. A. George, S. Shivaraman,
  V. B. Shields, M. Chandrashekhar, J. Hwang, F. Rana, M. G. Spencer,
  C. S. Ruiz-Vargas, and J. Park, Appl. Phys. Lett. {\bf 96}, 081917
  (2010).

\bibitem{Ruzicka2010} B. A. Ruzicka, L. K. Werake, H. Zhao, S. Wang, and K. P. Loh,
  Appl. Phys. Lett. {\bf 96}, 173106 (2010).

\bibitem{Huang} L. Huang, G. V. Hartland, L. Chu, Luxmi, R. M. Feenstra,
  C. Lian, K. Tahy, and H. Xing, Nano Lett. {\bf 10}, 1308 (2010).

\bibitem{ShangACSNano}J. Shang, T. Yu, J. Lin, and G. G. Gurzadyan, ACS Nano {\bf 5}, 3278 (2011).

\bibitem{WinnerlPRL} S. Winnerl, M. Orlita, P. Plochocka, P. Kossacki,
  M. Potemski, T. Winzer, E. Malic, A. Knorr, M. Sprinkle, C. Berger, W. A. de
  Heer, H. Schneider, and M. Helm, Phys. Rev. Lett. {\bf 107}, 237401 (2011).


\bibitem{Hale} P. J. Hale, S. M. Hornett, J. Moger, D. W. Horsell, and
  E. Hendry, Phys. Rev. B {\bf 83}, 121404(R) (2011).


\bibitem{BreusingPRB83} M. Breusing, S. Kuehn, T. Winzer, E. Malic, F. Milde,
  N. Severin, J. P. Rabe, C. Ropers, A. Knorr, and T. Elsaesser, Phys. Rev. B
  {\bf 83}, 153410 (2011).

\bibitem{Dani2012} K. M. Dani, J. Lee, R. Sharma, A. D. Mohite, C. M. Galande, P. M. Ajayan,
A. M. Dattelbaum, H. Htoon, A. J. Taylor, and R. P. Prasankumar, Phys. Rev. B
{\bf 86}, 125403 (2012).

\bibitem{LiPRL108} T. Li, L. Luo, M. Hupalo, J. Zhang, M. C. Tringides, J. Schmalian, and
J. Wang, Phys. Rev. Lett. {\bf 108}, 167401 (2012).


\bibitem{BridaArXiv}D. Brida, A. Tomadin, C. Manzoni, Y. J. Kim, A. Lombardo,
  S. Milana, R. R. Nair, K. S. Novoselov, A. C. Ferrari, G. Cerullo, and
  M. Polini, arXiv:1209.5729. 

\bibitem{Tielrooij} K. J. Tielrooij, J. C. W. Song, S. A. Jensen, A. Centeno, A. Pesquera,
A. Zurutuza Elorza, M. Bonn, L. S. Levitov, and F. H. L. Koppens,
arXiv:1210.1205v1.

\bibitem{Sun} B. Y. Sun, Y. Zhou, and M. W. Wu, Phys. Rev. B {\bf 85}, 125413 (2012).

\bibitem{ButscherAPL91}S. Butscher, F. Milde, M. Hirtschulz, E. Mali\'{c}, and A. Knorr, Appl. Phys. Lett. {\bf 91}, 203103 (2007).

\bibitem{Winzer} T. Winzer, A. Knorr, and E. Malic, Nano Lett. {\bf 10},
  4839 (2010).

\bibitem{Kim} R. Kim, V. Perebeinos, and P. Avouris, Phys. Rev. B {\bf 84}, 075449 (2011).

\bibitem{Malic} E. Malic, T. Winzer, E. Bobkin, and A. Knorr, Phys. Rev. B {\bf
  84}, 205406 (2011).

\bibitem{WinzerPRB85} T. Winzer and E. Mali\'{c}, Phys. Rev. B {\bf 85}, 241404(R)
  (2012).

\bibitem{WinzerArX1209} T. Winzer, E. Mali\'{c}, and A. Knorr, arXiv:1209.4833.


\bibitem{TaniArxiv}S. Tani, F. Blanchard, and K. Tanaka, arXiv:1206.1392v1.

\bibitem{Satou} A. Satou, V. Ryzhii, Y. Kurita, and T. Otsuji,
  arXiv:1210.6704v2.

\bibitem{Sachdev} S. Sachdev, Phys. Rev. B {\bf 57}, 7157 (1998).

\bibitem{Fritz} L. Fritz, J. Schmalian, M. M\"{u}ller, and S. Sachdev, Phys. Rev. B
  {\bf 78}, 085416 (2008).

\bibitem{Private} E. Mali\'{c}, Private communication.

\bibitem{HwangPRB76} E. H. Hwang, B. Y.-K. Hu, and S. Das Sarma, Phys. Rev. B {\bf 76}, 115434 (2007).

\bibitem{TseaAPL93} W.-K. Tse, E. H. Hwang, and S. Das Sarma, Appl. Phys. Lett. {\bf
    93}, 023128 (2008).

\bibitem{yzhou} Y. Zhou and M. W. Wu, Phys. Rev. B {\bf 82}, 085304 (2010).

\bibitem{Wunsch} B. Wunsch, T. Stauber, F. Sols, and F Guinea, New J. Phys. {\bf 8}, 318 (2006).

\bibitem{Muller} M. M\"{u}ller, L. Fritz, and S. Sachdev, Phys. Rev. B {\bf 78}, 115406 (2008).

\bibitem{Haug} H. Haug and A.-P. Jauho, {\it Quantum Kinetics in
    Transport and Optics of Semiconductors} (Springer, Berlin, 1998).
\bibitem{wuReview} M. W. Wu, J. H. Jiang, and M. Q. Weng,
  Phys. Rep. {\bf 493}, 61 (2010).
\bibitem{XFWang} X.-F. Wang and T. Chakraborty, Phys. Rev. B {\bf 75},
  033408 (2007).

\bibitem{Hwang} E. H. Hwang and S. Das Sarma, Phys. Rev. B {\bf 75},
  205418 (2007).

\bibitem{Ramezanali} M. R. Ramezanali, M. M. Vazifeh, R. Asgari,
  M. Polini, and A. H. MacDonald, J. Phys. A: Math. Theor. {\bf 42},
  214015 (2009).



\bibitem{DiVincenzoPRB} D. P. DiVincenzo and E. J. Mele, Phys. Rev. B {\bf 29}, 1685 (1984).


\bibitem{PeresPRB} N. M. R. Peres, F. Guinea, and A. H. Castro Neto,
  Phys. Rev. B {\bf 73}, 125411 (2006).

\bibitem{Stroucken2011} T. Stroucken, J. H. Gr\"onqvist, and S. W. Koch,
  Phys. Rev. B {\bf 84}, 205445 (2011).

\bibitem{Malic2006} E. Mali\'c, M. Hirtschulz, F. Milde, A. Knorr, and S. Reich,
  Phys. Rev. B {\bf 74}, 195431 (2006).

\bibitem{PZhang} P. Zhang and M. W. Wu, Phys. Rev. B {\bf 84}, 045304 (2011).

\bibitem{PZhang2} P. Zhang and M. W. Wu, New J. Phys. {\bf 14}, 033015 (2012).

\bibitem{ValencebandHF} The influence of the full valence band is subtracted 
in  our calculation following Ref.~\onlinecite{Haug} (Chap. 16).

\bibitem{Hwang2} E. H. Hwang, S. Adam, and S. Das Sarma,
  Phys. Rev. Lett. {\bf 98}, 186806 (2007).




\bibitem{Adam} S. Adam, E. H. Hwang, V. M. Galitski, and S. Das Sarma,
  Proc. Natl. Acad. Sci. U.S.A. {\bf 104}, 18392 (2007).

\bibitem{Adam2} S. Adam and S. Das Sarma, Solid State Commun. {\bf
    146}, 356 (2008). 
\bibitem{Fratini} S. Fratini and F. Guinea, Phys. Rev. B {\bf 77},
  195415 (2008).


\bibitem{HaugKoch} H. Haug and S. W. Koch, {\it Quantum Theory of the Optical and Electronic Properties of Semiconductors} (World Scientific, Singapore, 2004 4th ed.).


\bibitem{Weng} M. Q. Weng, M. W. Wu, and L. Jiang, Phys. Rev. B {\bf
    69}, 245320 (2004).


\bibitem{Rana} F. Rana, IEEE Trans. Nanotechnol. {\bf 7}, 91 (2008).

\bibitem{Dawlaty2} J. M. Dawlaty, S. Shivaraman, J. Strait, P. George,
  M. Chandrashekhar, F. Rana, M. G. Spencer, D. Veksler, and Y. Chen,
  Appl. Phys. Lett. {\bf 93}, 131905 (2008).

\bibitem{Scharf} B. Scharf, V. Perebeinos, J. Fabian, and P. Avouris,
  Phys. Rev. B {\bf 87}, 035414 (2013).

\bibitem{anisotropic} When the distribution is isotropic,
    Eq.~(\ref{conductivity}) can be simplified to the one given in
    Ref.~\onlinecite{Dawlaty} [$\sigma_{\omega_{\rm pr}}(t)=-{e^2}[{f}_{{\bf
        k_\omega},1}(t)-{f}_{{\bf k_\omega},-1}(t)]/4$]. 

\bibitem{Piscanec} S. Piscanec, M. Lazzeri, F. Mauri, A. C. Ferrari,
  and J. Robertson, Phys. Rev. Lett. {\bf 93}, 185503 (2004).

\bibitem{Lazzeri} M. Lazzeri, S. Piscanec, F. Mauri, A. C. Ferrari, and
  J. Robertson, Phys. Rev. Lett. {\bf 95}, 236802 (2005).

\bibitem{Hwang3} E. H. Hwang and S. Das Sarma, Phys. Rev. B {\bf 77}, 115449 (2008).

\bibitem{Chen} J.-H. Chen, C. Jang, S. Xiao, M. Ishigami, and M. S. Fuhrer,
  Nat. Nanotechnol. {\bf 3}, 206 (2008).


\bibitem{Ghosh} G. Ghosh, Opt. Commun. {\bf 163}, 95 (1999).

\bibitem{Perebeinos} V. Perebeinos and P. Avouris, Phys. Rev. B {\bf
    81}, 195442 (2010).

\bibitem{Novikov} D. S. Novikov, Phys. Rev. B {\bf 76}, 245435 (2007).

\bibitem{Levinshtein} M. E. Levinshtein, S. L. Rumyantsev, and M. S. Shur, {\it
    Properties of Advanced Semiconductor Materials: GaN, AlN, InN, BN, SiC, SiGe}
  (John Wiley \& Sons, New York, 2001).


\bibitem{explainTaupp} According to the fitting of the experimental data here
  (Sec. IIIA2), $\tau_{\rm pp}$ is 2.8~ps.
  
\bibitem{Dorgan2010} V. E. Dorgan, M.-H. Bae, and E. Pop, Appl. Phys. Lett. {\bf
    97}, 082112 (2010).

\bibitem{Pi2010} K. Pi, W. Han, K. M. McCreary, A. G. Swartz, Y. Li, and
  R. K. Kawakami, Phys. Rev. Lett. {\bf 104}, 187201 (2010).

\bibitem{Lifshitz} E. M. Lifshitz and L. P. Pitaevskii, {\it Physical Kinetics} (Pergamon,
London, 1981); G. E. Uhlenbeck, G. W. Ford, and E. W. Montroll, {\it Lectures in
  Statistical Mechanics} (American Mathematical Society, Providence, 1963),
Chap. IV; V. F. Gantmakher and Y. B. Levinson, {\it Carrier Scattering in Metals and Semiconductors} (North-Holland, Amsterdam, 1987), Chap. 6.


\bibitem{Zhang2010} P. Zhang and M. W. Wu, Europhys. Lett. {\bf 92}, 47009 (2010).

\bibitem{Latil} S. Latil, V. Meunier, and L. Henrard, Phys. Rev. B {\bf 76},
  201402(R) (2007).
\bibitem{Varchon} F. Varchon, P. Mallet, L. Magaud, and J.-Y. Veuillen,
Phys. Rev. B {\bf 77}, 165415 (2008).



\bibitem{Fischetti2001} M. V. Fischetti, D. A. Neumayer, and E. A. Cartier, J.
Appl. Phys. {\bf 90}, 4587 (2001).


\bibitem{Harris1995} G. Harris, {\it Properties of Silicon Carbide} (INSPEC, Institution of Electrical Engineers, London, UK, 1995).


\bibitem{Ishikawa} K. Ishikawa and T. Ando, J. Phys. Soc. Jpn. {\bf 75}, 084713 (2006).


\bibitem{Carrillo2003} J. A. Carrillo, I. M. Gamba, A. Majorana, and C. W. Shu,
  J. Comput. Phys. {\bf 184}, 498 (2003).

\end{thebibliography}
\end{document}